\documentclass[twocolumn,showpacs,superscriptaddress]{revtex4}

\usepackage{amsmath,amssymb}
\usepackage{graphicx}
\usepackage{mathrsfs}
\usepackage{color}

\begin{document}

\title{L\'{e}vy walk dynamics in an external harmonic potential}

\author{Pengbo Xu}
\affiliation{School of Mathematics and Statistics, Gansu Key Laboratory of
Applied Mathematics and Complex Systems, Lanzhou University, Lanzhou 730000,
P.R. China}
\author{Tian Zhou}
\affiliation{School of Mathematics and Statistics, Gansu Key Laboratory of
Applied Mathematics and Complex Systems, Lanzhou University, Lanzhou 730000,
P.R. China}
\author{Ralf Metzler}
\affiliation{Institute for Physics \& Astronomy, University of Potsdam,
Karl-Liebknecht-St 24/25, 14476 Potsdam, Germany}
\author{Weihua Deng}
\affiliation{School of Mathematics and Statistics, Gansu Key Laboratory of
Applied Mathematics and Complex Systems, Lanzhou University, Lanzhou 730000,
P.R. China}

\begin{abstract}
L{\'e}vy walks (LWs) are spatiotemporally coupled random-walk processes describing
superdiffusive heat conduction in solids, propagation of light in disordered optical
materials, motion of molecular motors in living cells, or motion of animals, humans,
robots, and viruses. We here investigate a key feature of LWs, their response to an
external harmonic potential. In this generic setting for confined motion we
demonstrate that LWs equilibrate exponentially and may assume a bimodal stationary
distribution. We also show that the stationary distribution has a horizontal slope
next to a reflecting boundary placed at the origin, in contrast to correlated
superdiffusive processes. Our results generalize LWs to confining forces and
settle some long-standing puzzles around LWs.
\end{abstract}

\maketitle

Anomalous diffusion with mean squared displacement (MSD) $\langle x^2(t)\rangle
\simeq t^{\alpha}$, whose anomalous diffusion exponent differs from the value
$\alpha=1$ of Brownian motion, are ubiquitously observed in a wide range of
systems \cite{bouchaud,pccp,franosch}. Subdiffusion with $0<\alpha<1$
occurs in amorphous semiconductors \cite{scher}, artificially crowded liquids
\cite{weiss}, lipid bilayer membranes \cite{schwille,weigel,ilpo}, cytoplasm
of biological cells \cite{lene,tabei}, or in hydrology \cite{grl}. Superdiffusion
with $\alpha>1$ is observed in active systems such as molecular motor transport
in cells \cite{roberts,christine,seisenhuber} or in turbulence \cite{igorturb}.
One of the central stochastic models for both regimes of anomalous diffusion is
the continuous time random walk (CTRW), based on the two identically distributed
random variables of the waiting times $\tau$ in between any two jumps and the
single jump lengths $x$ \cite{montroll,scher,bouchaud,report}. In the hydrodynamic
limit uncoupled CTRW processes in an external potential can be conveniently
described in terms of time- and/or space-fractional Fokker-Planck equations
\cite{mebakla,report,fogedby,jespersen}.

Superdiffusion is often modeled by L{\'e}vy flights (LFs), CTRWs with exponential
waiting time probability density function (PDF) and power-law jump length PDF
$\lambda(x)\simeq|x|^{-1-\mu}$ ($0<\mu<2$) \cite{fogedby}. The scale-free nature
of $\lambda(x)$ translates into a diverging MSD, but transport can be characterized
in terms of fractional order moments $\langle|x|^\kappa\rangle^{2/\kappa}\simeq t^{
2/\mu}$ \cite{report}. Due to their fractal, clustering motion pattern LFs are often
used as efficient random search mechanisms, e.g., for foraging animals
\cite{ghandi,vladimir}. In harmonic external potentials LFs have a stationary state
yet diverging MSD \cite{jespersen}. In steeper than harmonic potentials LFs assume
multimodal stationary PDFs with finite MSD but diverging higher-order moments
\cite{chechkin}.

A physically more pleasing CTRW concept for superdiffusion are L{\'e}vy walks
(LWs), based on a spatiotemporal coupling of jump lengths and waiting times with
a finite propagation speed and finite MSD \cite{wang,zaburdaev}. This property
makes them ideal candidates for the description of anomalous heat transport
\cite{dhar}, transport in Lorentz-like gases \cite{eliyossi}, and light propagation
in disordered optical media \cite{light}. LWs were shown to be efficient search
strategies \cite{michael,sims}, consistent with their first-hitting time properties
\cite{vladimir1}, and may emerge from deterministic nonlinear systems near a critical
point \cite{abe}. Indeed, LWs are observed in molecular-motor motion \cite{jae},
spreading of cancer cells \cite{cancer}, human hunter-gatherer foraging \cite{hunt},
pedestrian movement \cite{pedestrian}, and in optimized robotic search \cite{robot}.
LWs underly human movement patterns \cite{brockmann} and were identified in the
COVID-19 pandemic propagation \cite{havlin}.

LWs are ``ultraweakly" non-ergodic and fulfill generalized fluctuation-dissipation
relations \cite{froemberg,godec} and are related to infinite densities
\cite{rebenstok}. For constant external drift LWs are described by a fractional
material derivative \cite{sokolov}, for arbitrary external potentials LWs follow
a generalized Kramers-Fokker-Planck equations \cite{friedrich}. The latter is hard
to solve for concrete problems, as the Fourier-Laplace technique cannot be applied
due to the spatiotemporal coupling. Here we report an explicit solution of LWs in
a physically important harmonic potential. Answering some puzzles in LW theory, we
demonstrate that the PDF relaxes \emph{exponentially\/} to a \emph{stationary
limit\/} with a plateau value of the MSD that is independent of the exact
formulation of the LW.  We moreover demonstrate that the stationary PDF is
\emph{bimodal} in a wide parameter range. When the process approaches a regular
random walk a monomodal stationary PDF is restored. The PDF is also shown to have
a horizontal asymptote in the presence of a reflecting boundary placed at the
origin.

The scenario with harmonic confinement is relevant for molecular motors tethered
to a center (e.g., an intersection between microtubules in a cell, or a cargo that
is stuck in the cytoskeleton) by a flexible linker. Similarly the LW could be a
motor attached to a cargo that is in the harmonic potential of an optical tweezer.
On a macroscopic scale, the harmonic confinement models the restriction on animal
and human motion imposed by the ``territory'' (home range, quarantine restrictions,
etc.).

\emph{LWs in harmonic external potential.} We first consider a random walker
with mass $M$ and position $x_t$ at time $t$ in the harmonic potential $V(x_t)=
\frac{\gamma}{2}x_t^2$ with constant $\gamma>0$. Let $x_t$ be the final position
of each step of the LW. According to \cite{zaburdaev} we consider the starting
velocity of each step to be $\pm v_0$ ($v_0>0$) with probability of $1/2$
for left and right ($-$ or $+$). This picture is similar to a skater, whose
initial push is always identical. As the skater's speed diminishes while
gliding, in the course of a step the LW's velocity is changed by the potential.
Denote $t_i$ ($i=1,2,\ldots,n$) the time when the $i$th renewal event just
finishes and assume that the duration $\tau=t_i-t_{i-1}$ between two renewal
events obeys the density $\phi(\tau)$. Then
\begin{equation}
\label{sec1_eq1}
Md^2 x_{t_i+\tau'}/d{\tau'}^2=-\gamma x_{t_i+\tau'},
\end{equation}
governs the dynamics between the $i$th and $(i+1)$th renewals, for $t_i\le t$
and $\tau'\in(0,\{t_{i+1}\wedge t\}-t_i]$, for initial position $x_{t_i}$ and
velocity $dx_{t_i+\tau'}/d\tau'|_{\tau'=0}=\pm v_0$. The solution of
(\ref{sec1_eq1}) is $x_{t_i+\tau'}=x_{t_i}\cos(\omega\tau')\pm\frac{v_0}{\omega}
\sin(\omega\tau')$, where $\omega=\sqrt{\gamma/M}$. According to the theory of
LWs \cite{zaburdaev},
\begin{equation}
\label{sec1_eq2}
\begin{split}
q(x_{t'},t')=&\int_{-\infty}^{\infty}dx_{t'-\tau}\int_0^tq(x_{t'-\tau},t'-\tau)\\
&\times\upsilon(x_{t'-\tau},x_{t'},\tau)\phi(\tau) d t_{\tau}+p_0(x_{t'})\delta(t')
\end{split}
\end{equation}
determines the PDF $q(x_{t'},t')$ that the renewal event finishes at time $t'<t$
and the particle arrives at position $x_{t'}$. $p_0(x)=\delta(x)$ is the initial
PDF and $\upsilon(x,y,\tau)=\frac{1}{2}\delta(y- x\cos(\omega\tau)+\frac{v_0}{\omega}
\sin(\omega \tau))+\frac{1}{2}\delta(y- x\cos(\omega\tau)-\frac{v_0}{\omega}\sin(
\omega \tau))$. With the property of the delta function we rewrite (\ref{sec1_eq2})
as
\begin{eqnarray}
\nonumber
&q(x_{t'},t')-p_0(x)\delta(t')=(1/2)\int_0^{t'}\phi(\tau)d\tau/|\cos(\omega\tau)|\\
&\times[q(x_{t'}^+,t'-\tau)+q(x_{t'}^-,t'-\tau)],
\label{fb_q}
\end{eqnarray}
where $x_{t'}^\pm=(x_{t'}/\cos(\omega \tau))\pm(v_0/\omega)\tan(\omega \tau)$.
The PDF $p(x,t)$ to find the particle at $x$ at time $t$ then satisfies
\begin{equation}
\label{fb_p}
p(x,t)=\int_{-\infty}^{\infty}\int_0^tq(x_{t-\tau},t-\tau)\upsilon(x_{t-\tau},x,\tau)
\Psi(\tau)d\tau dx_{t-\tau},
\end{equation}
where $\Psi(\tau)=\int_{\tau}^{\infty}\phi(\tau')d\tau'$. With $\upsilon(x_{t
-\tau},x,\tau)$ we get
\begin{equation}\label{fb_p_2}
p(x,t)=\frac{1}{2}\int_0^t\frac{\Psi(\tau)d\tau}{|\cos(\omega\tau)|}
\left[q(x_{t'}^+,t'-\tau)+q(x_{t'}^-,t'-\tau)\right].
\end{equation}
We express $p(x,t)$ in terms of Hermite polynomials $H_n(x)$ \cite{xu2018}.
These are orthogonal to each other over $(-\infty,\infty)$ with weight $\exp(-x^2)$
\cite{hermit_intro}. We respectively take
\begin{equation}
\label{fb_assume_form_q}
\{q(x,t),p(x,t)\}=\sum_{n=0}^{\infty}H_n(x)e^{-x^2}\{T_n(t),\tilde{T}_n(t)\},
\end{equation}
where the eigenfunctions $T_n(t)$, $\tilde{T}_n(t)$ are determined in \cite{supp}.
The PDF composed of Eqs.~(S5) and (S6) allows us to calculate the
stationary PDF and statistical quantities.

Consider now $\hat{T}_m(s)$, $\hat{\tilde{T}}_m(s)$ for odd $m$.
When $m=1$ from (S5) we deduce that $\hat{T}_1(s)=\mathscr{L}\{\cos(
\omega\tau)\phi(\tau)\}\hat{T}_1(s)$, implying $\hat{T}_1(s)=0$. Analogously, $\hat{
\tilde{T}}_1(s)=0$ from (S6). By induction, for every odd $m$,
$T_m(t)=\tilde{T}_m(t)=0$. Therefore in \eqref{fb_assume_form_q} only even terms are
left, and thus $q(x,t)$ and $p(x,t)$ are even functions, reflecting the symmetry of
the problem. The $m$th moment is given by $\langle x^m(t)\rangle=i^m\frac{d^m}{d
k^m}\bar{p}(k,t)\big|_{k=0}$, with the Fourier transform $\bar{p}(k,t)=\int_{-\infty}
^{\infty}e^{-i k x}p(x,t)d_x=\sum_{n=0}^{\infty}\sqrt{\pi}(-i k)^{n}e^{-k^2/4}\tilde{
T}_n(t)$. The Laplace transform of the MSD is $\langle x^2(s)\rangle=\frac{\sqrt{
\pi}}{2}\hat{\tilde{T}}_0(s)+2\sqrt{\pi}\hat{\tilde{T}}_2(s)$, where $\hat{\tilde{T}}
_0(s)$ and $\hat{\tilde{T}}_2(s)$ can be obtained from (S6) and
(S6). With $\hat{\tilde{T}}_0(s)=(\sqrt{\pi}s)^{-1}$, we have
the normalization $\int_{-\infty}^{\infty}p(x,t)dx=\bar{p}(k=0,t)=\sqrt{\pi}\tilde{
T}_0(t)=1$.

For the MSD we obtain $\hat{T}_2(s)$, $\hat{\tilde{T}}_2$(s) from (S5), (S6) for
specific $\phi(\tau)$. For the exponential $\phi(\tau)=\beta e^{-\beta\tau}$, we
get $\hat{\tilde{T}}_2(s)=\frac{2 v_0^2-\omega^2}{4\sqrt{\pi}\omega^2 s}$. At long
$t$ (small $s$) the asymptotic behavior of the MSD is given by the constant
\begin{equation}
\label{MSD_fb}
\langle x^2(t)\rangle\sim v_0^2/\omega^2.
\end{equation}
For uniform $\phi(\tau)=\frac{1}{T}\mathbf{1}_{[0,T]}(\tau)$ on $[0,T]$ with period
$T=2\pi/\omega$ ($\mathbf{1}_{[0,T]}(\tau)$ is the indicator function) as well as
for the asymptotic power-law $\phi(\tau)=\alpha/(1+\tau)^{1+\alpha}$ ($\alpha>0$)
we find the same plateau \eqref{MSD_fb}. Thus LWs in a harmonic potential always
localize asymptotically and the plateau value only depends on the stiffness of the
potential as well as the speed $v_0$ and mass of the particle. The form of $\phi(
\tau)$ has no influence on the plateau \eqref{MSD_fb} and the sufficiently fast
decay of $p^{\mathrm{st}}(x)$ at $|x|\to\infty$ (see also below).

\begin{figure}
\includegraphics[width=4cm]{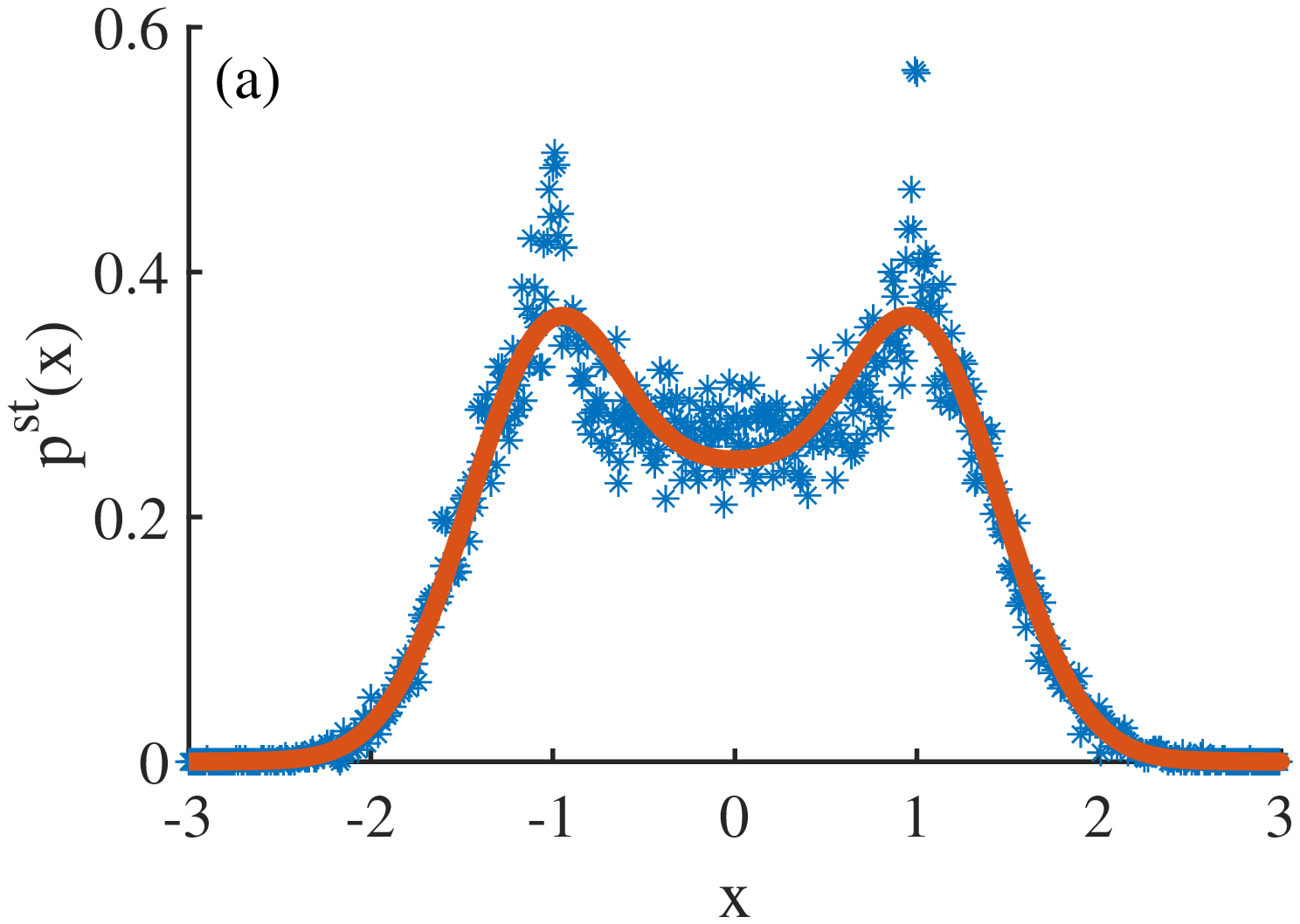}
\includegraphics[width=4cm]{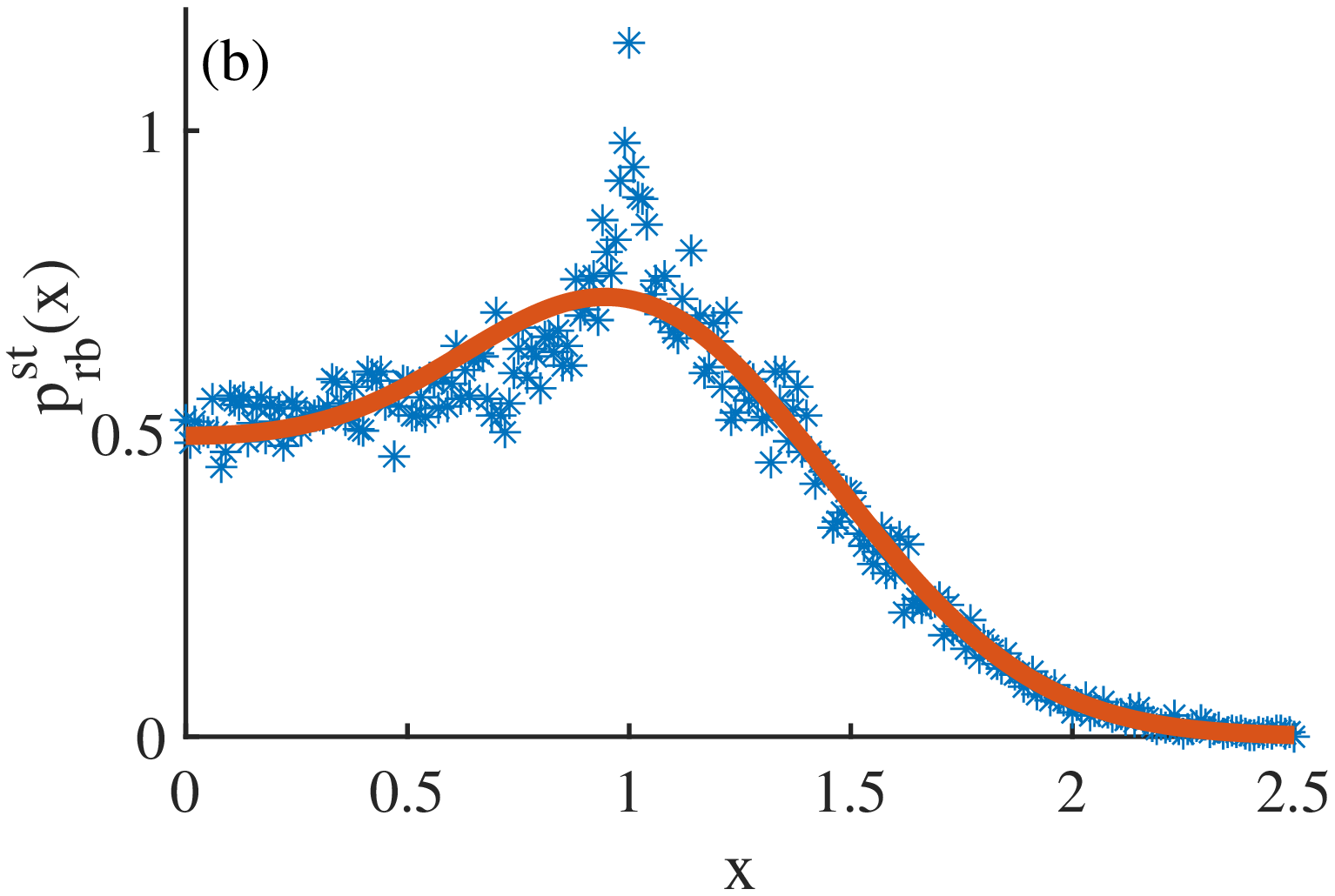}
\caption{Stationary PDF of LW in harmonic potential for $\phi(\tau)=\beta e^{
-\beta\tau}$, $v_0=\omega=\beta=1$. Stars: simulations from $10^4$ realizations. (a)
No boundaries. Line: approximate theoretical result for
$N=13$ terms and simulation time $t=10^4$. (b) Reflecting boundary condition
at $x=0$, simulation time $t=10^3$. Line: approximate result $\lim_{
t\to\infty}\sum_{n=0}^{12}e^{-x^2}H_n\tilde{T}_n(t)$.}
\label{fig1}
\end{figure}

The stationary PDF follows from the final value theorem of the Laplace transform,
$p^{\mathrm{st}}(x)=\lim_{t\to\infty}p(x,t)=\lim_{s\to0}s\hat{p}(x,s)=\lim_{s\to0}
\sum_{n=0}^{\infty}H_{2n}(x)e^{-x^2}s\hat{\tilde{T}}_{2n}(s)$, and $\hat{\tilde{T}}
_{2n}(s)$ is given by (S5), (S6). For explicit calculations we truncate the series
after $N$ terms, to obtain the approximate stationary PDF for sufficiently large
$N$. We choose $\phi(\tau)=e^{-\tau}$ and $v_0/\omega=1$. For $N=13$ we find the
approximate stationary PDF in \cite{supp}, as shown in Fig.~\ref{fig1}a. Despite
the potential minimum at the origin, $p^{\mathrm{st}}(x)$ is distinctly bimodal
with maximum at $|x|\approx v_0/\omega$. Physically, the peaks emerge due to the fact
that each jump starting at the origin actually points away from $x=0$. We would
thus expect that for sufficiently large $v_0$ and appropriate systems parameters
the bimodality occurs. Note that similar effects are indeed known from LFs: an
LF in a harmonic potential is stationary and monomodal \cite{jespersen}, yet in
steeper than harmonic potentials LFs are bimodal \cite{chechkin}. We now further
explore $p^{\mathrm{st}}(x)$.

\begin{figure}
\includegraphics[width=4.0cm]{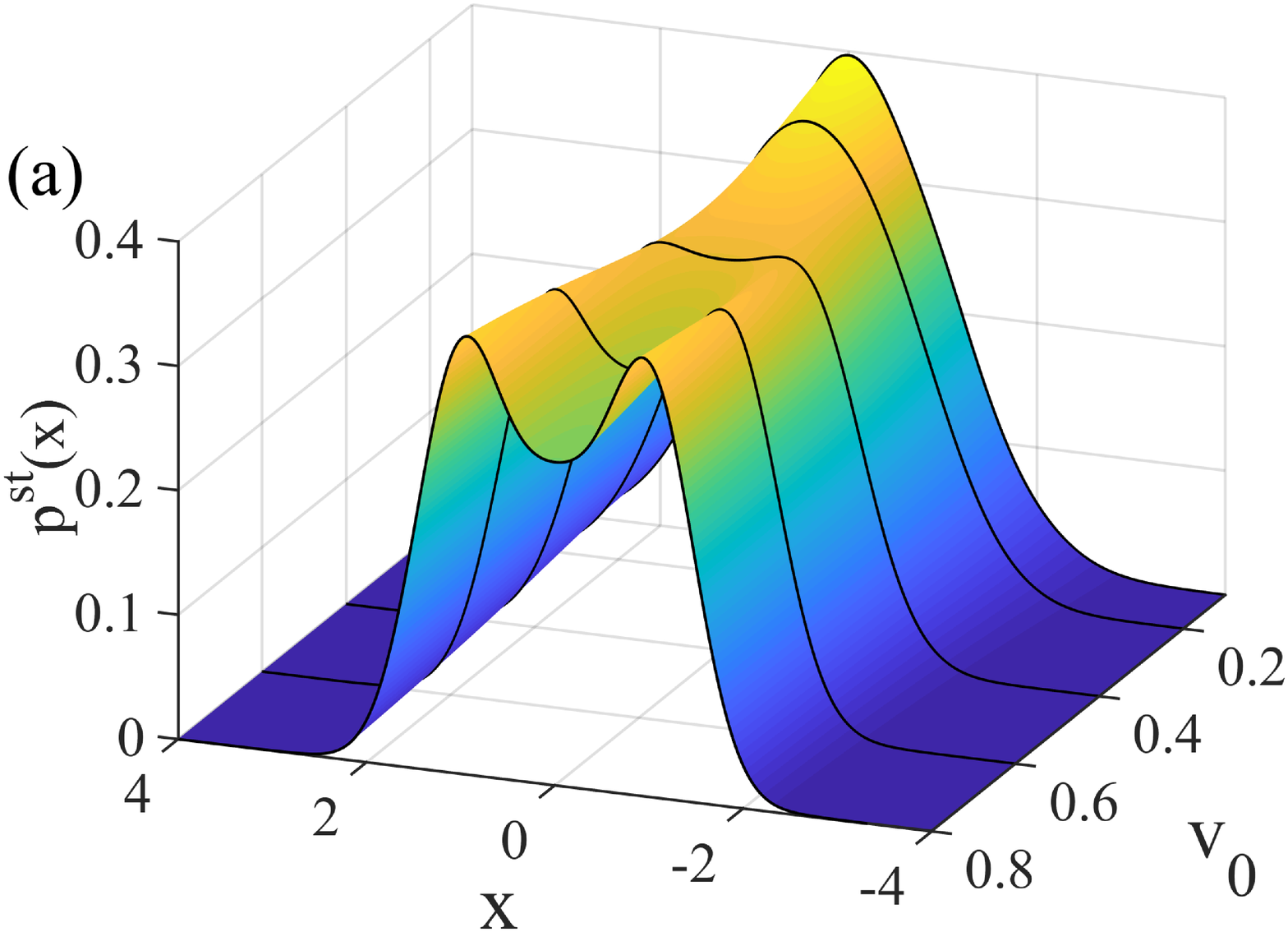}
\includegraphics[width=4.0cm]{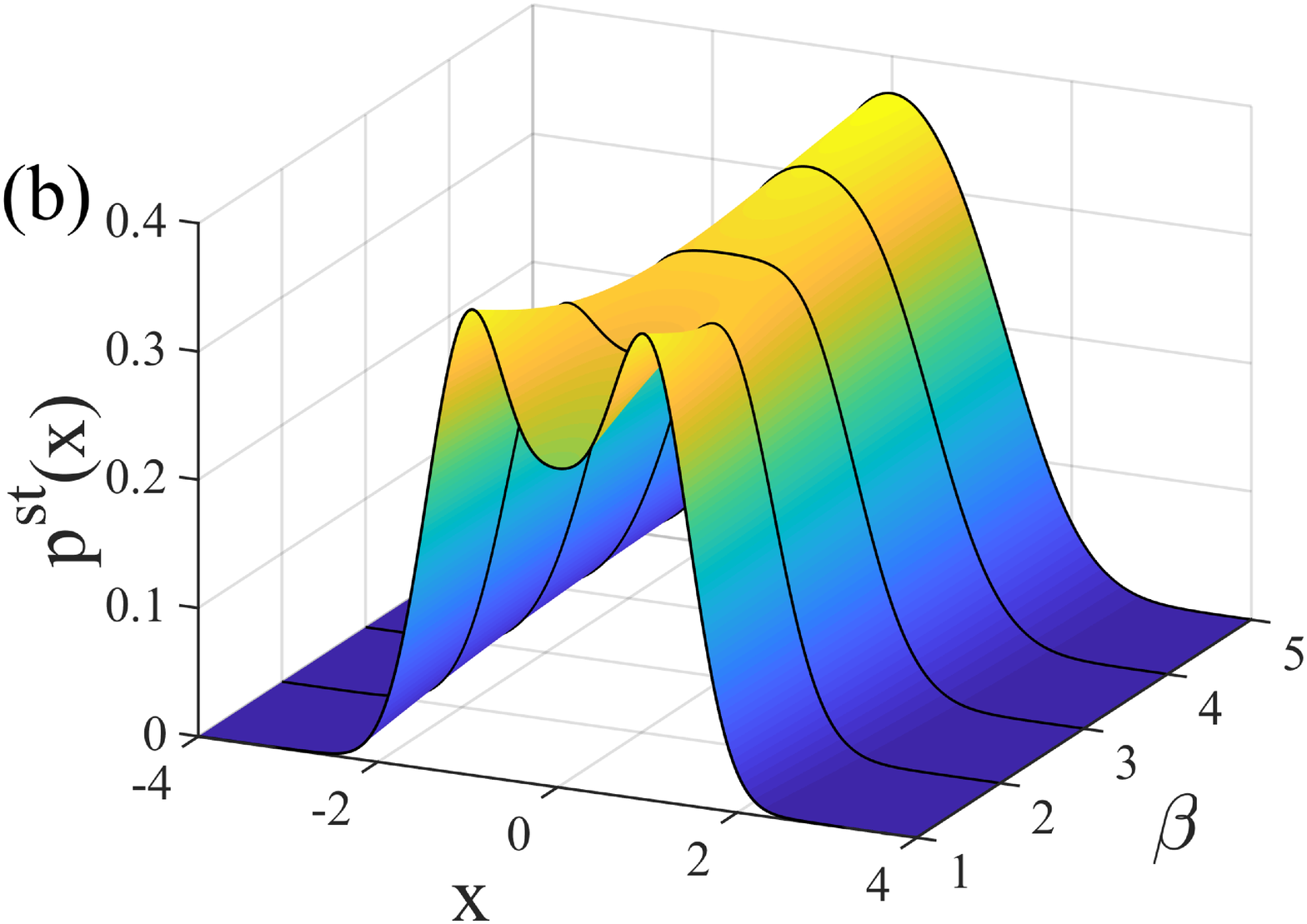}
\caption{Stationary PDF \cite{supp} for $\phi(\tau)=\beta e^{-\beta\tau}$ with
$\beta=1$ and varying $v_0$, and $v_0=1$ and changing $\beta$ with $v_0/\omega=1$
fixed. In both cases a monomodal-to-bimodal crossover occurs.}
\label{fig4}
\end{figure}

As shown in Fig.~\ref{fig4} for exponential $\phi(\tau)$ \cite{supp} the bimodality
of $p^{\mathrm{st}}(x)$ indeed depends on the exact parameters. Once $v_0$ is
small or $\beta$ becomes large, i.e., when the LW approaches the limit of a
regular random walk, monomodality is restored. As shown in Fig.~S1 similar
behaviors occur for uniform and power-law forms of $\phi(\tau)$. The crossover
can in fact be quite delicate; see Fig.~S1.

\begin{figure}
\includegraphics[height=3.0cm]{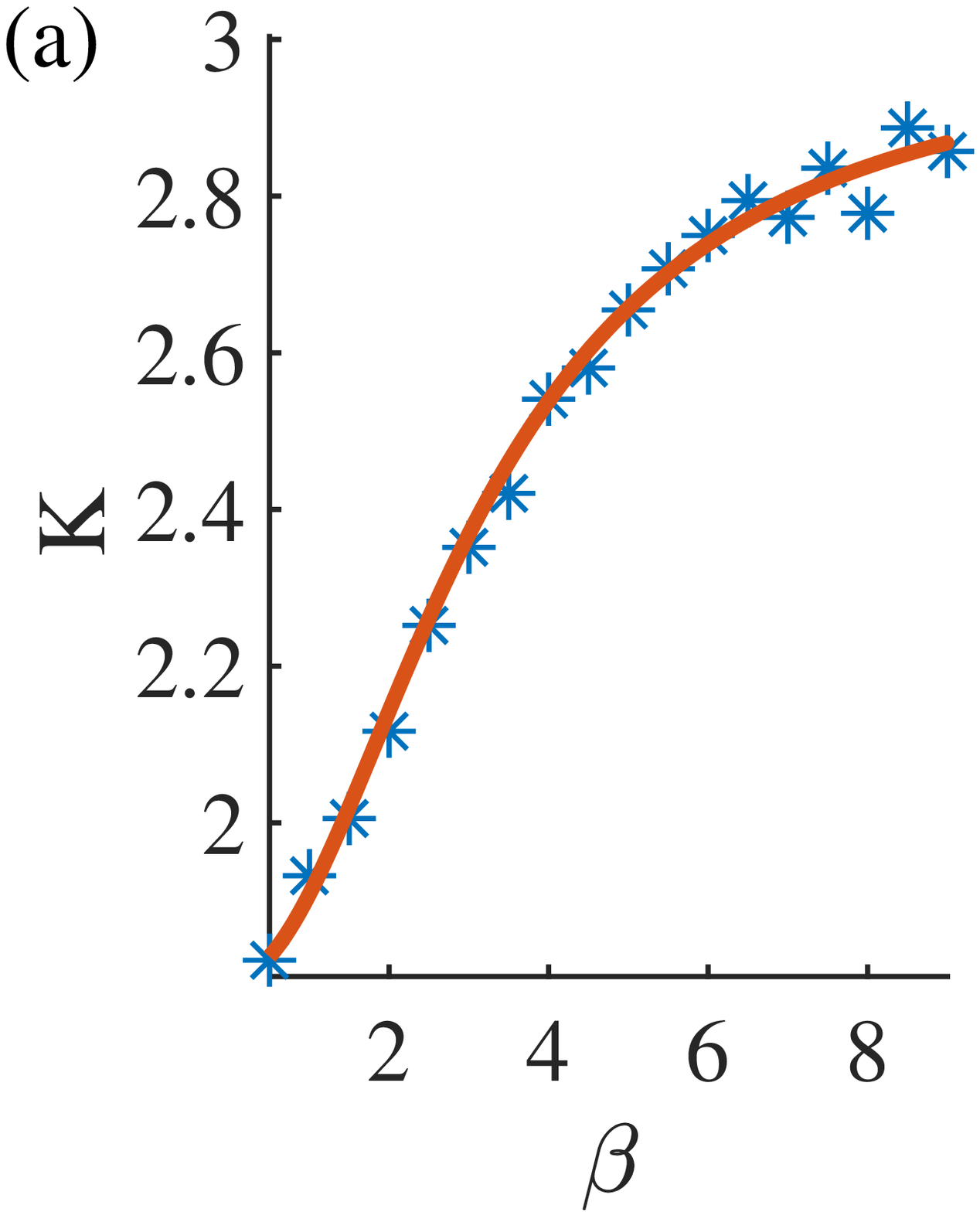}
\includegraphics[height=3.0cm]{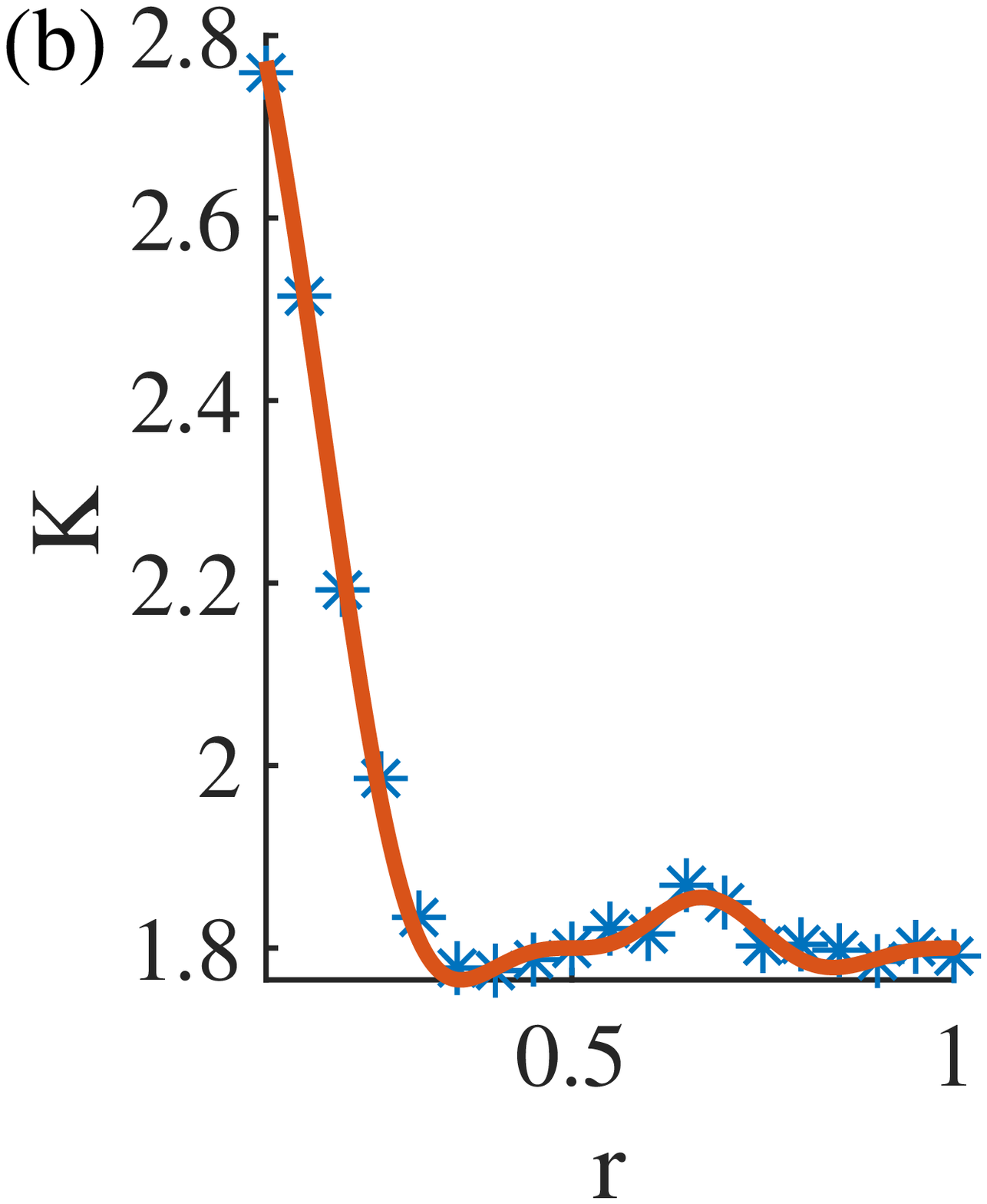}
\includegraphics[height=3.0cm]{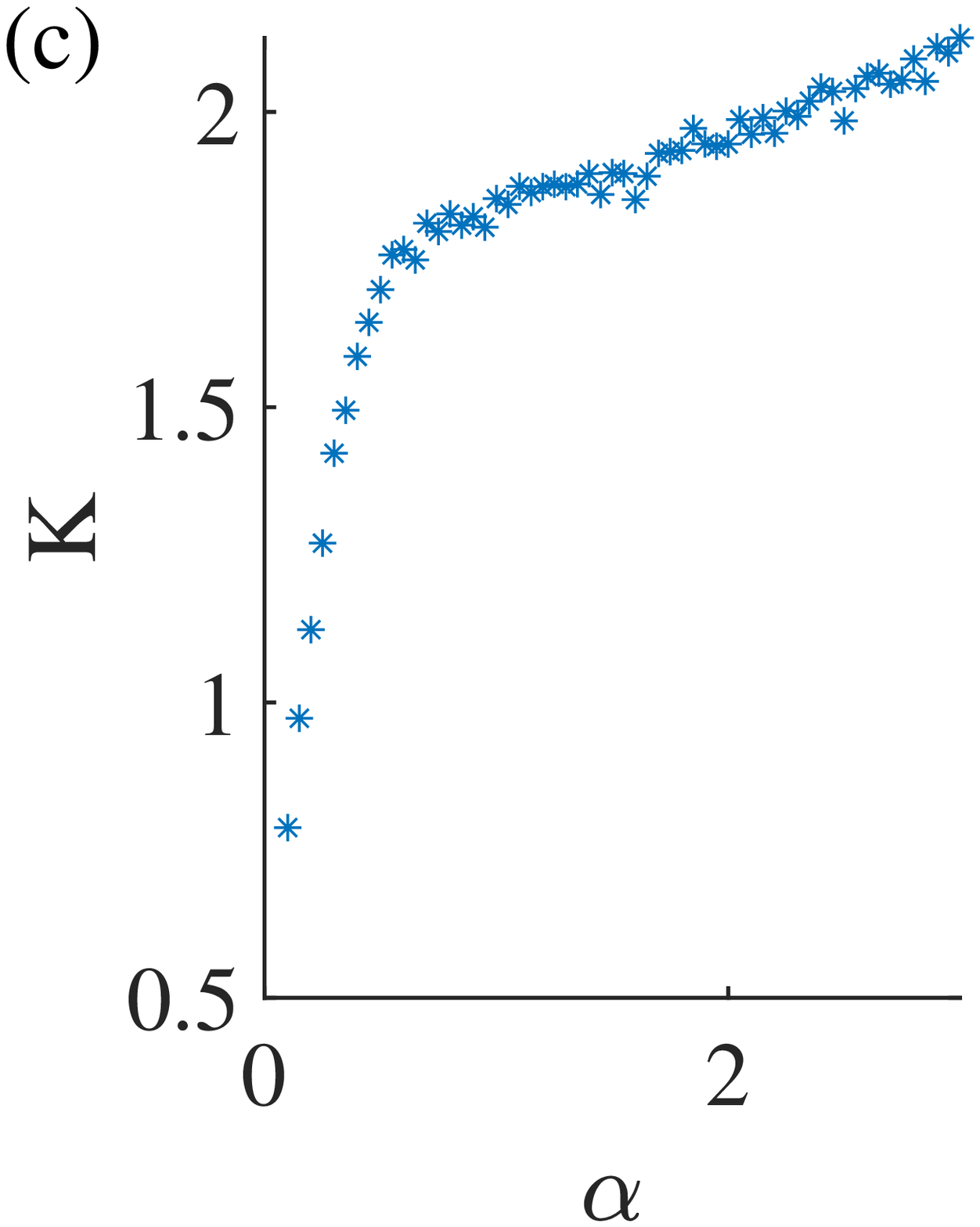}
\caption{Stationary value of the kurtosis $K$ for three $\phi(\tau)$ for
$v_0=\omega=1$, averaged over $5\times10^4$ trajectories and simulation
time $t=10^4$: (a) $\phi(\tau)=\beta e^{-\beta\tau}$ as function of $\beta$;
(b) $\phi(\tau)=\mathbf{1}_{[0,2\pi r/\omega]}(\tau)$ versus $r$; (c) $\phi(
\tau)=\alpha/(1+\tau)^{1+\alpha}$ versus $\alpha$. Values below 3 indicate
that the PDF is platykurtic.}
\label{fig2}
\end{figure}

The tails of the stationary PDF are characterized by the kurtosis $K=\langle x^4(t)
\rangle/\langle x^2(t)\rangle^2$. When $\phi(\tau) =\beta e^{-\beta\tau}$, Eqs.~(S5)
and (S6) lead to $\langle\hat{x}^4(s)\rangle=3\sqrt{\pi}\hat{\tilde{T}}_0/4+6\sqrt{
\pi}\hat{\tilde{T}}_2+24\sqrt{\pi}\hat{\tilde{T}}_4\sim[3v_0^4(\beta^2+6\omega^2)]/
[s\omega^2(\beta^2+10\omega^2)]$, i.e.
\begin{equation}
\label{kurt}
K=[3(\lambda^2+6\omega^2)]/(\lambda^2+10 \omega^2).
\end{equation}
This form is verified by simulations in Fig.~\ref{fig2}a. We note that, in contrast
to the MSD, the kurtosis depends on the shape of $\phi(\tau)$. For small inverse
time scales $\beta$, the $K$ values show that the PDF is platykurtic and converges
to the Gaussian value $K=3$ for large $\beta$. In this limit we expect the LW to
converge to a normal random walk, for which the PDF is Gaussian in a harmonic
potential. Analogous behavior is found for uniform $\phi(\tau)=\mathbf{1}_{[0,2\pi
r/\omega]}(\tau)$ \cite{supp}, Fig.~\ref{fig2}b. For small interval $r$, a Gaussian
emerges, as $K\sim3-2.4\pi^2r^2$. For an asymptotic power-law form $\phi(\tau)=
\alpha/(1+\tau)^{1+\alpha}$ simulations show that $K$ assumes platykurtic values
even for $\alpha>2$ (Fig.~\ref{fig2}c).

\begin{figure}
\includegraphics[width=4cm]{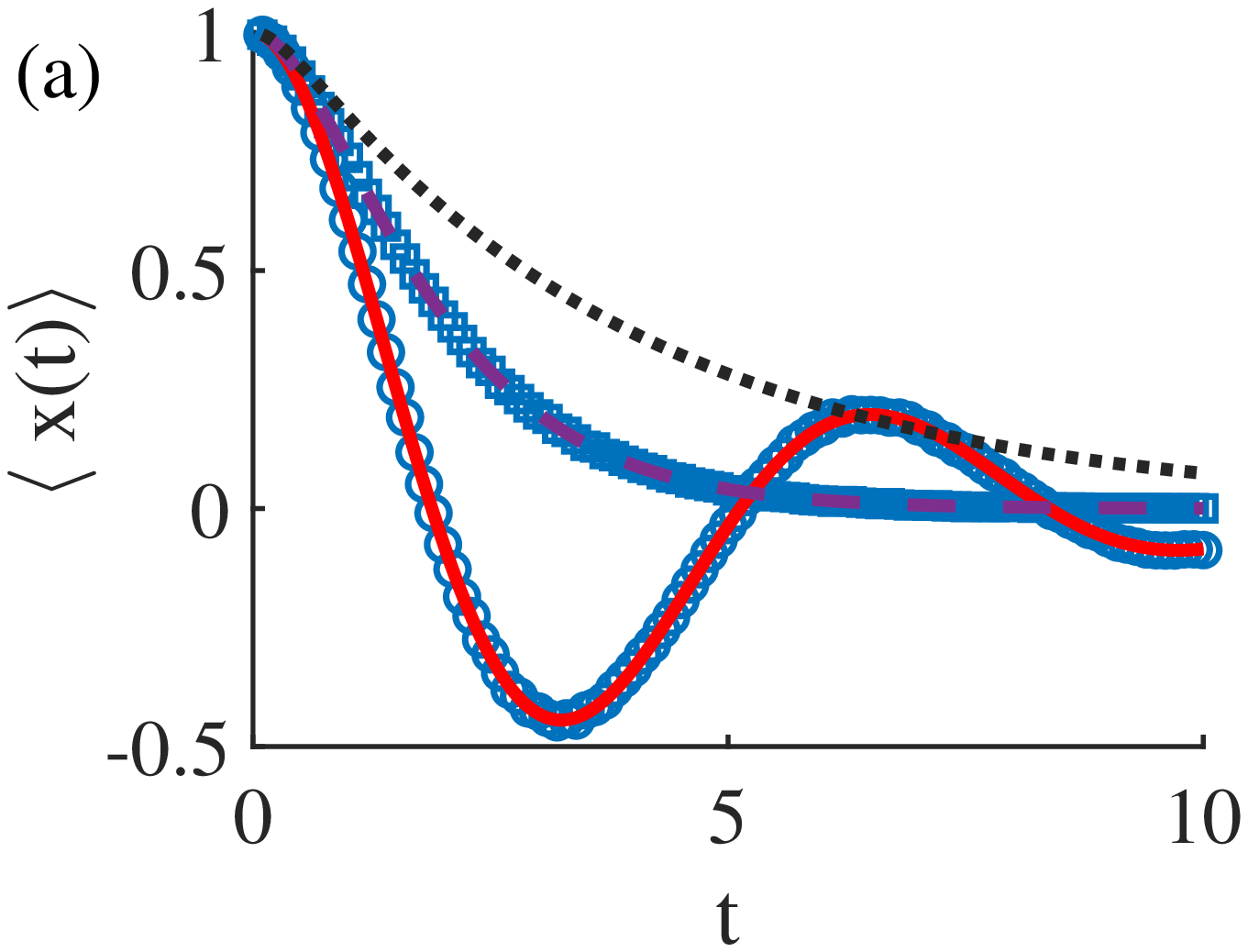}
\includegraphics[width=4cm]{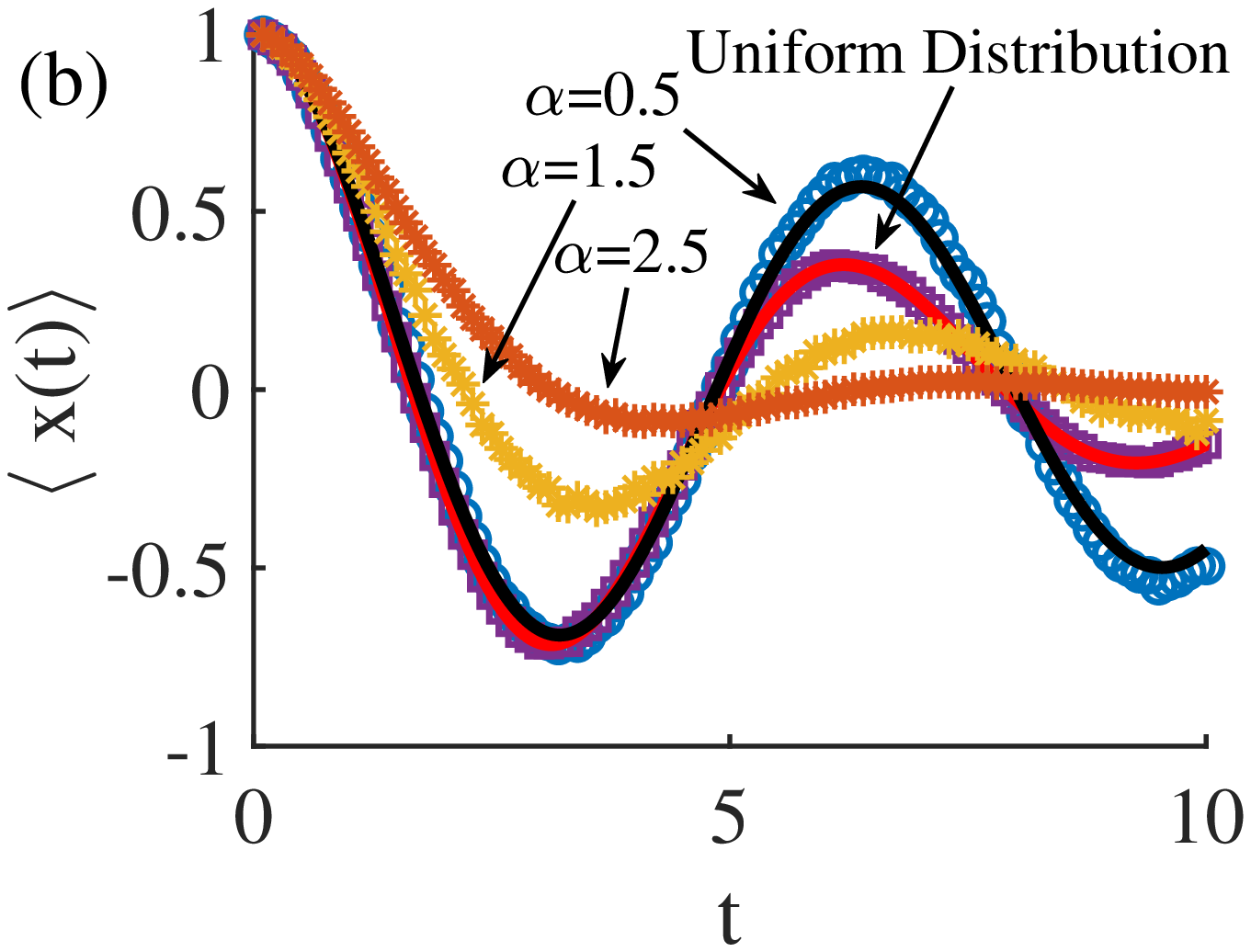}
\caption{Simulations results of the relaxation dynamics of the first moment
$\langle x(t)\rangle$ from $10^4$ realizations each, for $v_0=\omega=x_0=1$. (a)
exponential waiting time PDF $\phi(\tau)=\beta e^{-\beta}$ with $\beta=0.5$
(circles) and $\beta=2$ (squares). The full, dashed, and dotted (with $\beta=4$)
lines represent the theoretical results. (b) power-law and uniform densities
on $[0,2\pi]$, the lines are from numerical Laplace inversion. Note the oscillatory
behavior.}
\label{fig3}
\end{figure}

\emph{Relaxation dynamics.} We now discuss the relaxation of the LW particle in
the harmonic potential with initial position $x_0\neq0$, i.e., $p_0(x)=\delta(
x-x_0)$ for different forms of the waiting time $\phi(\tau)$. The mean position
is obtained as $\langle x(t)\rangle=\sqrt{\pi}\tilde{T}_1(t)$, where $\tilde{T}_1
(t)$ is given through
\begin{equation}
\label{fb_TT1_LT}
\hat{\tilde{T}}_1(s)=\frac{x_0\mathscr{L}\{\cos(\omega\tau)\Psi(\tau)\}}{\sqrt{
\pi}(1-\mathscr{L}\{\cos(\omega\tau)\phi(\tau)\})}.
\end{equation}
With $\mathscr{L}\{\cos(\omega t)f(t)\}=\frac{1}{2}(\hat{f}(s+i\omega)+\hat{f}(s-i
\omega))$ we get
\begin{equation}
\label{fb_TT1_rw}
\hat{\tilde{T}}_1(s)=\frac{x_0}{\sqrt{\pi}}\frac{\frac{1-\hat{\phi}(s-i\omega)}{s-i
\omega}+\frac{1-\hat{\phi}(s+i\omega)}{s+i\omega}}{2-\hat{\phi}(s-i\omega)-\hat{
\phi}(s+i\omega)}.
\end{equation}
For an exponential $\phi(\tau)$ we find $\langle\hat{x}(s)\rangle=\frac{x_0(s+
\beta)}{s^2+\omega^2+\beta s}$, i.e.,
\begin{equation}
\label{average_time_exp}
\begin{split}
&\hspace*{-1.6cm}\langle x(t)\rangle=x_0\exp(-[\beta+\sqrt{\beta^2-4\omega^2}]t/2)\\
&\hspace*{-1.2cm}\times\frac{1}{2}\left[(e^{\sqrt{\beta^2-4\omega^2}t}-1)/\sqrt{1-4
\omega^2/\beta^2}\right.\\
&\hspace*{-1.2cm}+\left.(e^{\sqrt{\beta^2-4 \omega^2}t}+1)\right];
\end{split}
\end{equation}
see Fig.~\ref{fig3}a.
Thus, the relaxation of the initial position is exponential to leading order.
For uniform $\phi(\tau)$ on $[0,T]$ with Laplace transform $\hat{\phi}(s)=(1
-e^{-T s})/(Ts)$,
\begin{equation}
\label{average_LT_uni}
\begin{split}
&\hspace*{-0.4cm}\langle\hat{x}(s)\rangle=\Big[e^{sT}(-s^2+s^3T+\omega^2+sT
\omega^2)+(s^2-\omega^2)\\
&\hspace*{-0.4cm}\times\cos(\omega T)-2s\omega\sin(\omega T)\Big]\Big/\Big[(s^2+
\omega^2)\\
&\hspace*{-0.4cm}\times(e^{sT}(-s+s^2T+T\omega^2)+s \cos(\omega T)-\omega\sin(\omega
T))\Big],
\end{split}
\end{equation}
which we analyze numerically in Fig.~\ref{fig3}b. The case of a
power-law form for $\phi(\tau)$ can only be solved numerically, after plugging
the asymptotic form $\hat{\phi}(s)\sim1-s^{\alpha}$ into \eqref{fb_TT1_rw}. The
resulting behavior is shown in Fig.~\ref{fig3}c.

In Fig.~\ref{fig3} we note the difference in the initial decay rate and
the final approach to zero. Curves with higher initial decay appear to converge
more slowly due to the apparent oscillations. Their existence reminds inertia
effects known from classical oscillators. In the present model they are likely
to the initial non-equilibrated speed $v_0$ for each jump. However, the exact
value of $v_0$ has no influence on the average displacement as seen from
\eqref{fb_TT1_rw}.

\emph{Reflecting boundary condition at $x=0$.} We now consider LWs in a harmonic
potential with a reflecting boundary at $x=0$. On the random walk level when the
$i$th step begins at time $t_{i-1}$ at position $|x_{t_{i-1}}|$, and it then moves
to $x_{t_i}$ which may be negative. Then for the $(i+1)$th step
we take the absolute value of the end displacement of step $i$, $|x_{t_i}|$, to
be the starting position of step $i+1$. For the last step ($n$, such that $t_n
+\tau>t$), we also need the absolute value $|x_t|$ as the end point of the walk.
In order to solve this problem, we first construct an auxiliary process, whose
last step is $x_t$ instead of $|x_t|$.

The detailed derivations are found in \cite{supp}. Fig.~\ref{fig1}b shows the
reflected stationary PDF. Note that for the reflected
process the PDF $p_{\mathrm{rb}}(|x|,t)$ can be given through $p_{\mathrm{rb}}(
|x|,t)=p_{\mathrm{aux}}(|x|,t)+p_{\mathrm{aux}}(-|x|,t)$. The MSD is $\int_0^{
\infty}x^2p_{\mathrm{rb}}(x,t)dx=\int_{-\infty}^{\infty}x^2p_{\mathrm{aux}}(x,t)
dx$, which indicates that the reflected and auxiliary processes have the same MSD,
as expected from the applicable method of images. Consequently the asymptotic
value of the MSD is given by \eqref{MSD_fb}. We note that the horizontal shape of
the PDF next to the reflecting boundary is a consequence of the renewal character
of the CTRW process. For positively (negatively) correlated stochastic processes
an accretion or depletion of probability occurs at the boundary \cite{vojta}.

\emph{Conclusions.} We considered LWs in a generic external harmonic potential.
Apart from being experimentally relevant, our results answer the conceptual
question whether and how LWs equilibrate in soft confinement. Our analysis
shows that LWs under harmonic confinement equilibrate to a stationary PDF, that,
surprisingly, may be bimodal with peak locations $x=\pm v_0/\omega$. However,
the bimodality delicately depends on the model parameters. When the LW approaches
a regular random walk, monomodality is restored. For exponential and uniform
$\phi(\tau)$ we also demonstrated that the stationary PDF in these limits becomes
Gaussian. While the stationary value of the MSD is independent of the chosen form
of $\phi(\tau)$ and thus in all cases the tails of the stationary PDF always decay
sufficiently fast, higher order moments depend on $\phi(\tau)$. This was discussed
for the fourth-order moment entering the kurtosis $K$. Our results for $K$ show
that the stationary PDF is always platykurtic.

The bimodality of LWs in a harmonic external potential are similar to the known
results for spatiotemporally decoupled LFs. While LFs are monomodal in a harmonic
potential and have diverging MSD, in steeper-than-harmonic potentials LFs assume
bimodal stationary PDFs. The main difference is that the stationary PDF of LFs
always have a power-law asymptote and thus the kurtosis is either undefined or has
a leptokurtic value.

The relaxation dynamics, as discussed for the mean particle position, was studied
by analytics and numerics for the three scenarios of the waiting time
density $\phi(\tau)$. In particular, we observe characteristic, pseudo-inertial
oscillations reflecting the ``skater'' formulation of the LW process adopted here,
namely, that each step starts with a fixed initial speed $v_0$. The results are
analogous for the case of a reflecting boundary at the origin, for which we showed
that the PDF is horizontal at the boundary, in contrast to correlated processes.

Following recent results for the onset of superdiffusion in LWs and their
behavior in finite domains \cite{miron} our work fills another gap in the
description of these widely used spatiotemporally coupled random walks.

\acknowledgments

This work was supported by the National Natural Science Foundation of China,
grant 11671182. RM acknowledges the German Science Foundation (DFG), grant
ME 1535/7-1 and the Foundation for Polish Science (FNP) for a Humboldt Polish
Honorary Research Scholarship.

\clearpage
\newpage
\onecolumngrid

\renewcommand\theequation{S\arabic{equation}}
\renewcommand\thefigure{S\arabic{figure}}
\setcounter{figure}{0}

\begin{center}
\textbf{\large Supplementary material: L\'{e}vy walk dynamics in an external harmonic
potential}
\end{center}

\section{Auxiliary calculations for the eigenfunction expression of the
probability density function}

Starting with expression (6) of the main text we now
define $\langle f(ax+b),g(cx+d)\rangle=\int_{-\infty}^{\infty}f(ax+b)g(cx+d)e^{
-(ax+b)^2}dx$, which does not satisfy linearity. We insert $q(x,t)$ from
Eq.~(6) into Eq. (3), multiply by $H_m(x)$ ($m=0,1,\ldots$) on both sides,
and integrate $x$ over $(-\infty,\infty)$:
\begin{equation}
\label{hermit_fb_q}
\begin{split}
&\sum_{n=0}^{\infty}\langle H_n(x),H_m(x)\rangle T_n(t)-H_m(0)\delta(t)\\
&=\frac{1}{2}\sum_{n=0}^{\infty}\int_{0}^{t} d\tau\bigg[\langle H_n\Big(
\frac{x}{\cos(\omega \tau)}+\frac{v_0}{\omega}\tan(\omega\tau)\Big),H_m(x)\rangle\\
&+\langle H_n\Big(\frac{x}{\cos(\omega \tau)}-\frac{v_0}{\omega}\tan(\omega\tau)
\Big),H_m(x)\rangle\bigg]\frac{\phi(\tau)T_n(t-\tau)}{|\cos(\omega\tau)|}\\
&=\frac{1}{2}\sum_{n=0}^{\infty}\int_{0}^{t} d\tau\bigg[\langle H_n(y),H_m\Big(
\cos(\omega\tau) y -\frac{v_0}{\omega}\sin(\omega\tau)\Big)\rangle\\
&+\langle H_n(y),H_m\Big(\cos(\omega\tau) y +\frac{v_0}{\omega}\sin(\omega\tau)
\Big)\rangle\bigg]\phi(\tau)T_n(t-\tau).
\end{split}
\end{equation}
We invoke the properties of the Hermite polynomials \cite{hermit_11,hermit_intro1}
\begin{equation}
\langle H_m(x),H_n(x)\rangle=2^nn!\sqrt{\pi}\delta_{n,m}
\end{equation}
with the Kronecker $\delta_{n,m}$, and
\begin{eqnarray}
\label{hermit_prop_sum}
&&H_n(x+y)=\sum_{k=0}^{n}\bigg(\begin{array}{c}n\\k\end{array}
\bigg)H_k(x)(2y)^{n-k},\\
\label{hermit_prop_time}
&&H_n(\gamma x)=\sum_{i=0}^{\lfloor\frac{n}{2}\rfloor}\gamma^{n-2i}
(\gamma^2-1)^i\bigg(\begin{array}{c}n\\2i\end{array}\bigg)\frac{2i!}{i!}H_{n-2i}(x),
\end{eqnarray}
where $\lfloor\frac{n}{2}\rfloor$ is the biggest integer smaller than $\frac{n}{2}$.
Laplace transforming, $\hat{f}(s)=\mathscr{L}\{f(t)\}=\int_0^{\infty}e^{-st}f(t)dt$
yields
\begin{equation}
\label{fb_T_laplace}
\begin{split}
&\hspace*{-0.2cm}\hat{T}_m(s)-\frac{H_m(0)}{\sqrt{\pi}2^m m!}\\
&\hspace*{-0.4cm}=\sum_{k=0}^{m}\sum_{i=0}^{\lfloor\frac{k}{2}\rfloor}\frac{
2^{-2i-1}}{(m-k)!i!}\left(\frac{v_0}{\omega}\right)^{m-k} \big[(-1)^i+(-1)^{
m-k+i}\big]\\
&\hspace*{-0.4cm}\times\mathscr{L}\left\{\sin^{m-k+2i}(\omega\tau)\cos^{k-2i}
(\omega\tau)\phi(\tau)\right\}\hat{T}_{k-2i}(s).
\end{split}
\end{equation}
Similarly, we obtain the corresponding relation
\begin{eqnarray}
\nonumber
&\hat{\tilde{T}}_m(s)=\sum\limits_{k=0}^{m}\sum\limits_{i=0}^{\lfloor\frac{k}{2}
\rfloor}\frac{(-1)^i2^{-2i-1}}{(m-k)!i!}\left(\frac{v_0}{\omega}\right)^{m-k}\big[
1+(-1)^{m-k}\big]\\
&\hspace*{-0.8cm}\times\mathscr{L}\{\cos^{k-2i}(\omega\tau)\sin^{m-k+2i}(\omega
\tau)\Psi(\tau)\}\hat{T}_{k-2i}(s).
\label{fb_tildeT_laplace}
\end{eqnarray}
Eqs.~(\ref{fb_T_laplace}) and (\ref{fb_tildeT_laplace}) are used in the main text.

\section{Numerical simulations of stationary PDF}

Supplementing the behavior of the stationary LW-PDF $p^{\mathrm{st}}(x)$ for
exponential waiting time density shown in the main text, we here present analogous
results for other forms of $\phi(\tau)$ and variations of the associated parameters.

\begin{figure}
\includegraphics[width=8cm]{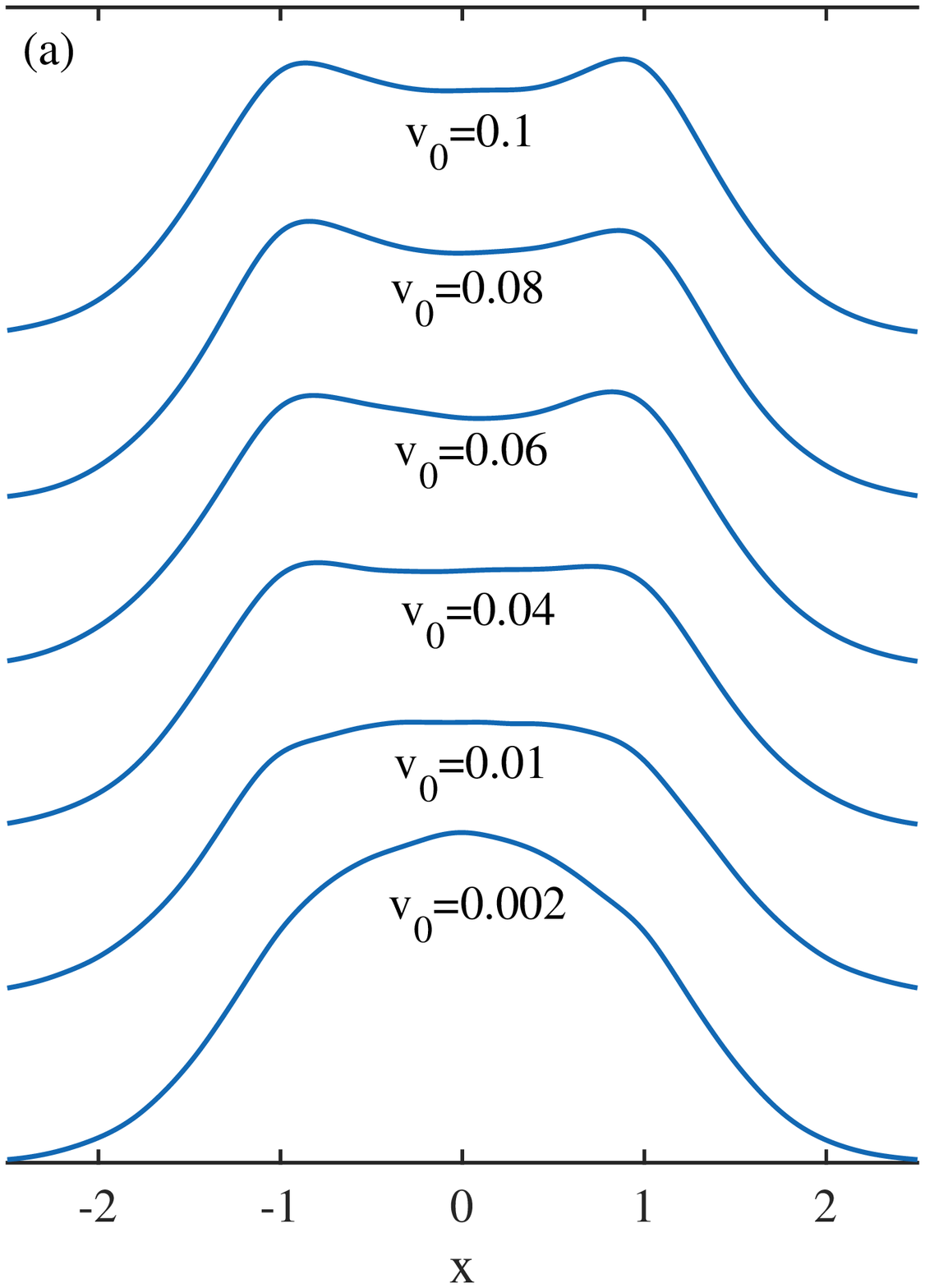}
\includegraphics[width=8cm]{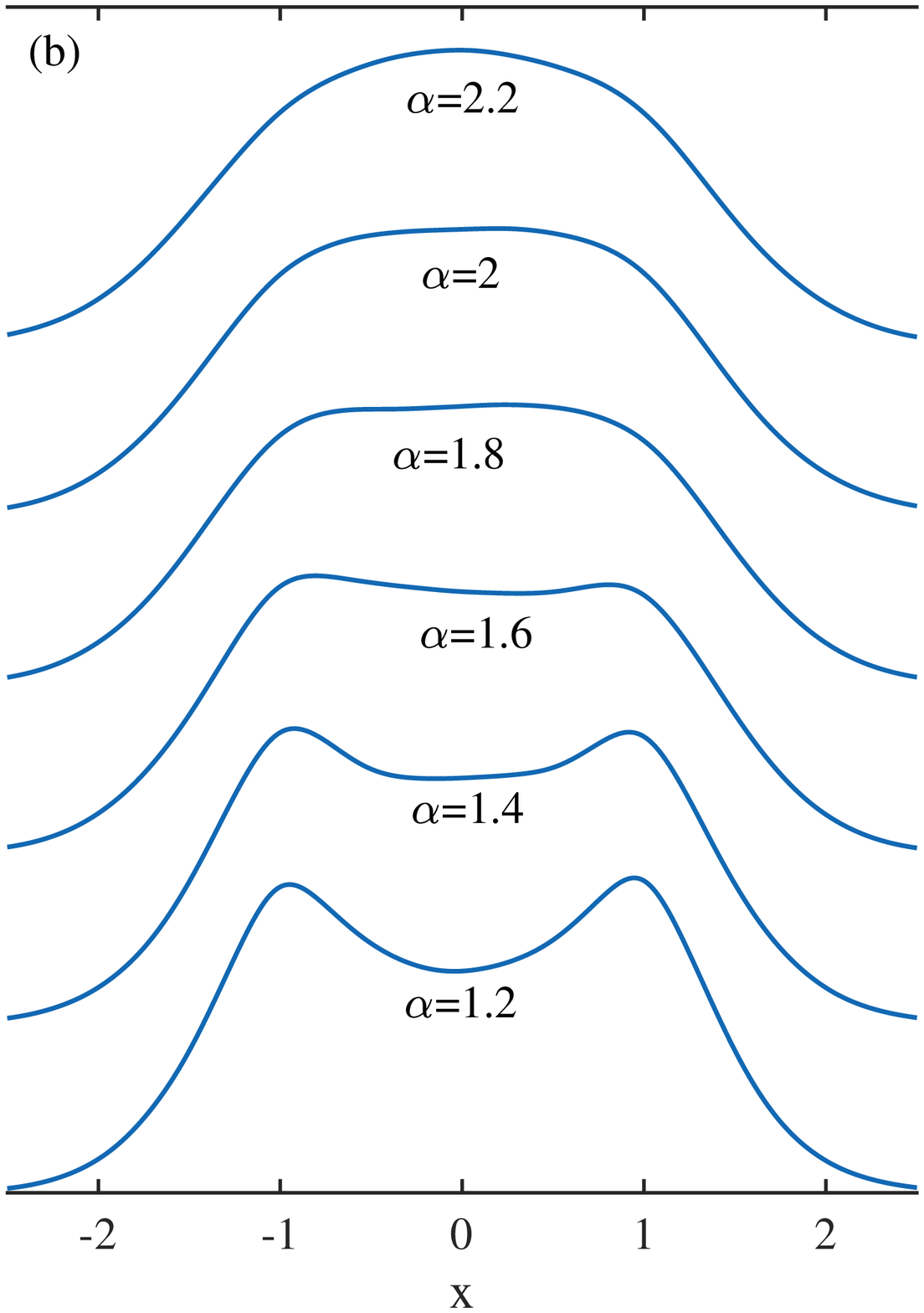}
\includegraphics[width=8cm]{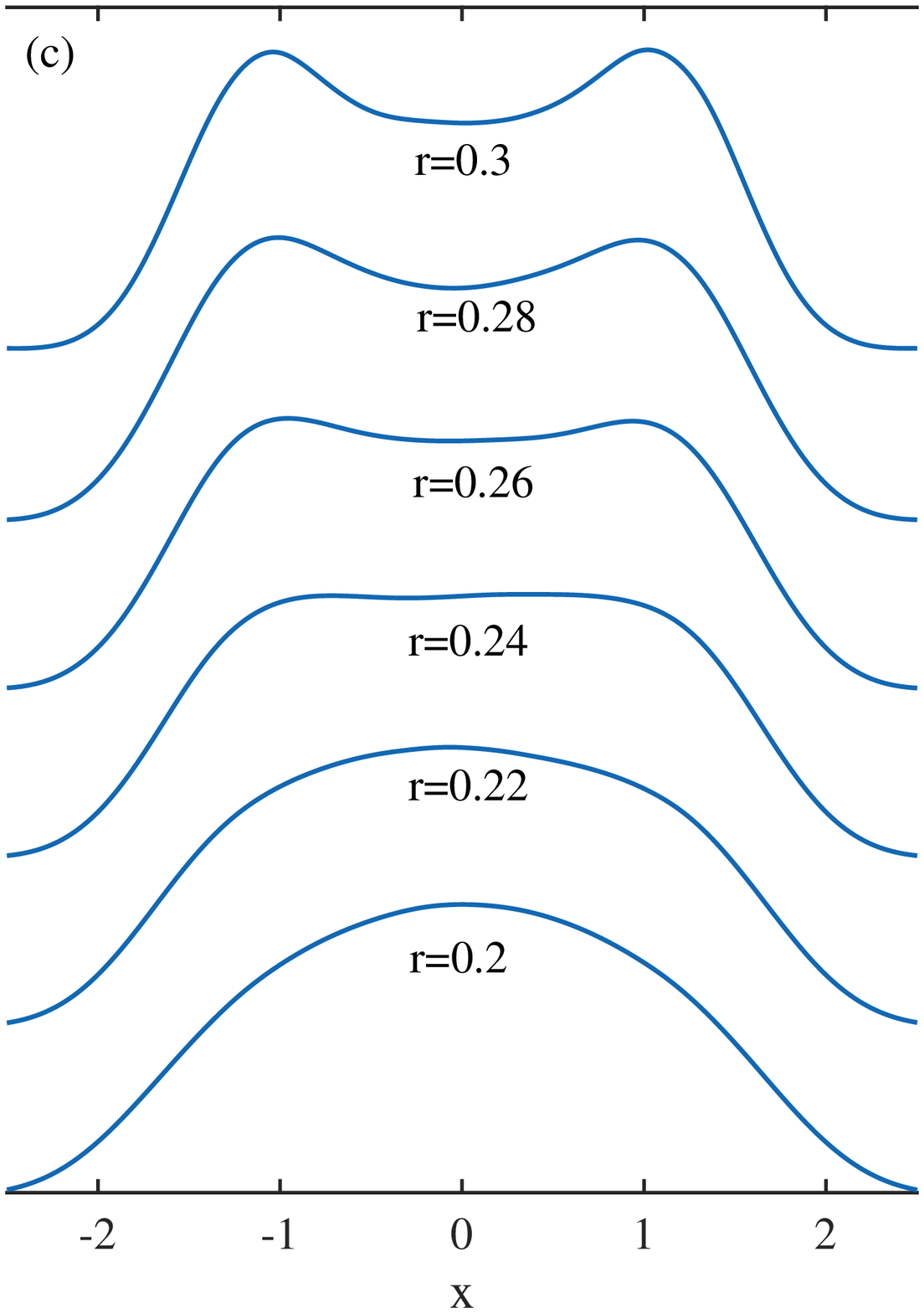}
\includegraphics[width=8cm]{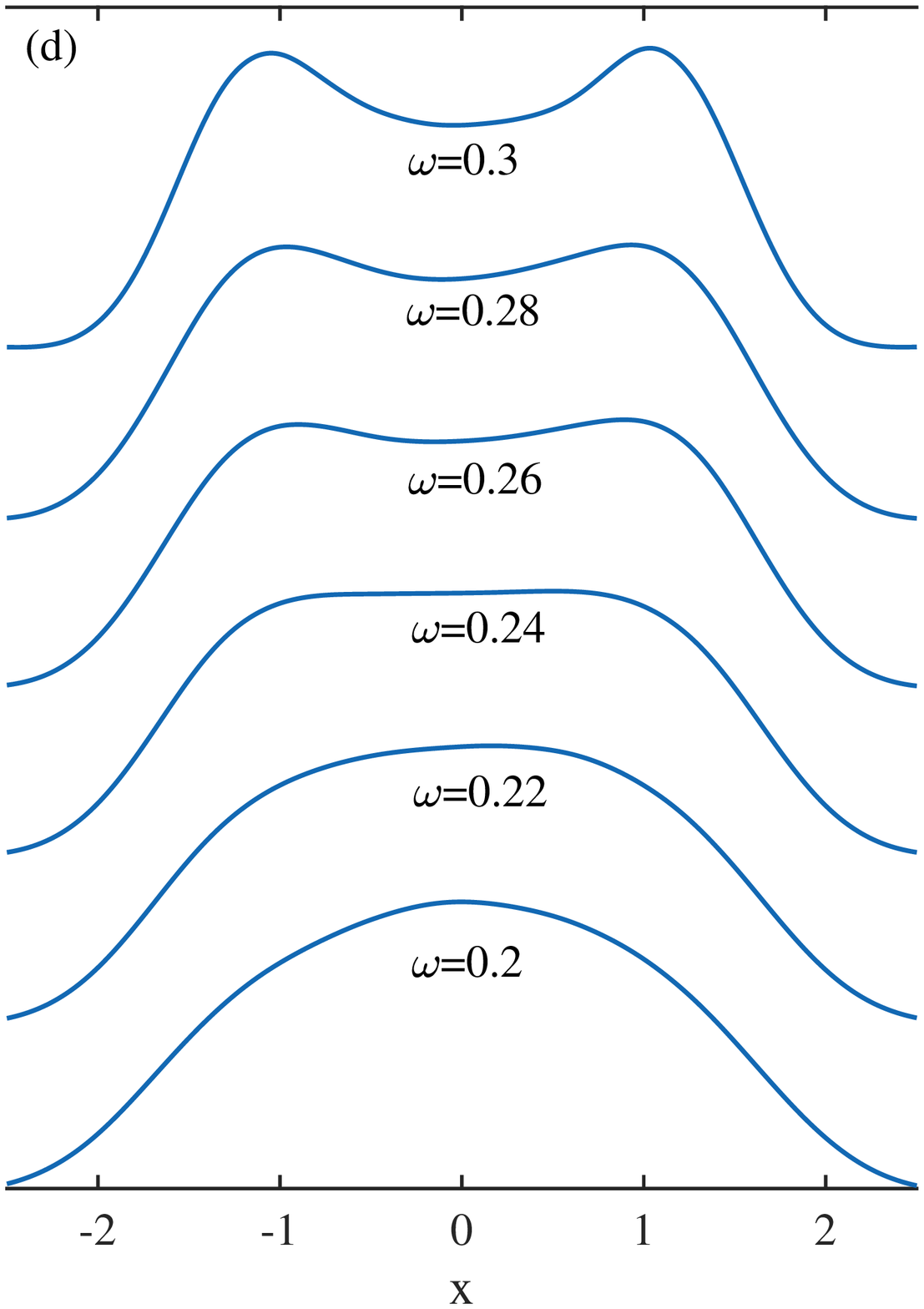}
\caption{Stationary PDFs from numerical simulations. For (a) and (b) the
asymptotic power-law waiting time PDF $\phi(\tau)=\alpha/(1+\tau)^{1+\alpha}$
was used with $v_0=\omega$. For (a) $\alpha=1.5$ and in (b) $v_0=0.1$. For (c)
and (d), the uniform waiting time PDF $\phi(\tau)=\mathbf{1}_{[0,2\pi r/\omega]}
(\tau)$ was used with $v_0=\omega$. In (c) $v_0=1$, and for (d) $v_0=\omega=r$,
so that the $\phi(\tau)$ are always same.}
\label{fig_fb_pst}
\end{figure}

First for the asymptotic power-law form $\phi(\tau)=\alpha/(1+\tau)^{1+\alpha}$,
Fig.~\ref{fig_fb_pst}a and \ref{fig_fb_pst}b show the effect of different speed
$v_0$ at the beginning of each jump, and of different powers $\alpha$. As we can
see, when $v_0$ and $\omega$ are sufficiently large, a bimodal stationary state
emerges. Similarly, when $\alpha$ is below the value 2 and thus the density $\phi
(\tau)$ abides to sufficiently long tails, bimodality is observed. Note that the
numerical accuracy we can achieve is not sufficient to numerically pin down the
crossover to monomodal behavior at exactly $\alpha=2$, but from the mathematical
nature of power-law distributions this assumption appears consequent.
Second, we consider the uniform density $\phi(\tau)=\mathbf{1}_{[0,2\pi r/\omega]}
(\tau)$ in Fig.~\ref{fig_fb_pst}c and d for different interval lengths $r$ and
$\omega$. When each of the two parameters becomes sufficiently small, monomodality
is restored. Note the delicate variation of the shapes with the second digit of
these parameters.

\section{Calculation of $\langle x^4(t)\rangle$ for LW in harmonic potential}

In the main text we provided the kurtsis for the case of an exponential waiting
time density $\phi(\tau)$. For the asymptotic power-law form for $\phi(\tau)$ we
resort to simulations. Here we calculate the fourth-order moment and the kurtosis
of an LW in a harmonic potential and without boundaries for the case of uniformly
distributed waiting time density $\phi(\tau)$ defined on $[0,2\pi r/\omega]$, $r>0$.
For this case, considering Eqs.~\eqref{fb_T_laplace} and \eqref{fb_tildeT_laplace}, the following
results can be obtained,
\begin{equation*}
\begin{split}
&\lim_{t\to\infty}T_0(t)=\frac{\omega}{\pi^{3/2}r};\\
&\lim_{t\to\infty}\tilde{T}_0(t)=\frac{1}{\sqrt{\pi}};\\
&\lim_{t\to\infty}T_2(t)=\frac{2v_0^2-\omega^2}{4\pi^{3/2}r\omega};\\
&\lim_{t\to\infty}\tilde{T}_2(t)=\frac{2v_0^2-\omega^2}{4\sqrt{\pi}\omega^2};\\
&\lim_{t\to\infty}T_4(t)=\frac{24\pi r(12v_0^4-20v_0^2\omega^2+5\omega^4)-
8(4v_0^4-12v_0^2\omega^2+3\omega^4)\sin(4\pi r)+(-20v_0^4+12v_0^2\omega^2-
3\omega^4)\sin(8\pi r)}{96\pi^{3/2}r\omega^3(40\pi r-8\sin(4\pi r)-\sin(8\pi r))};\\
&\lim_{t\to\infty}\tilde{T}_{4}(t)=\big[24\pi rv_0^4+1152\pi^3r^3v_0^4-1920\pi^3r^3v_0^2
\omega^2+480\pi^3r^3\omega^4-32\pi rv_0^4\cos(4\pi r)+8\pi rv_0^4\cos(8\pi r)+
5v_0^4\sin(4\pi r)\\
&~~~~~~~~~~~~-192\pi^2r^2v_0^4\sin(4\pi r)+384\pi^2r^2v_0^2\omega^2\sin(4\pi r)-
96\pi^2r^2\omega^4\sin(4\pi r)-4v_0^4\sin(8\pi r)-48\pi^2r^2v_0^4\sin(8\pi r)\\
&~~~~~~~~~~~~+48\pi^2r^2v_0^2\omega^2\sin(8\pi r)-12\pi^2r^2\omega^4\sin(8\pi r)+
v_0^4\sin(12\pi r)\big]\big/\big[384\pi^{5/2}r^2\omega^4(40\pi r-8\sin(4\pi r)-
\sin(8\pi r))\big].\\
\end{split}
\end{equation*}
Therefore,
\begin{equation*}
\begin{split}
\lim_{t\rightarrow\infty} \langle x^4(t)\rangle &=\frac{3 \sqrt{\pi}}{4} \lim_{t\rightarrow\infty} \tilde{T_0}(t) + 6 \sqrt{\pi} \lim_{t\rightarrow\infty} \tilde{T}_2(t) + 24 \sqrt{\pi} \lim_{t\rightarrow\infty} \tilde{T}_4(t) \\
& =v_0^4\big[24 \pi r + 1152 \pi^3 r^3 - 32 \pi r \cos(4 \pi r) +
8 \pi r \cos(8 \pi r) + 5 \sin(4 \pi r) -
192 \pi^2 r^2 \sin(4 \pi r) - 4 \sin(8 \pi r) \\
&~~~~-48 \pi^2 r^2 \sin(8 \pi r) +
\sin(12 \pi r)\big]\big/\big[16 \pi^2 r^2 \omega^4 (40 \pi r - 8 \sin(4 \pi r) -
\sin(8 \pi r))\big],\\
\lim_{t\to\infty}K&=\big[24\pi r+1152\pi^3r^3-32\pi r\cos(4\pi r)+8\pi r\cos(8\pi
r)+5\sin(4\pi r)-192\pi^2r^2\sin(4\pi r)-4\sin(8\pi r)\\
&~~~~-48\pi^2r^2\sin(8\pi r)+\sin(12\pi r)\big]\big/\big[16\pi^2r^2(40\pi r-8\sin(4
\pi r)-\sin(8\pi r))\big].
\end{split}
\end{equation*}
The series expansion of the kurtosis for small interval sizes $r$ then becomes
\begin{equation*}
K\sim3-\frac{12}{5}\pi^2r^2+\frac{88}{105}\pi^4r^4+\ldots,
\end{equation*}
quoted in the main text.

\section{Simplification of the governing equation for the PDF of velocity direction
change at a given position}
\label{supp_sec1}

In the presence of a \emph{reflecting boundary\/} we introduce the reflecting
condition as outlined in the main text. To construct the auxiliary function we
proceed as follows. Changing the initial condition of Eq.~(1) in the main text
from $x_{t_i}$ to $|x_{t_i}|$, we have $x_{t_i}=A\cos(\omega\tau'\pm\varphi_{
\pm v_0})$, where $A=\sqrt{x_{t_i}^2+v_0^2/\omega^2}$ and $\varphi_{\pm v_0}=
\arctan(\frac{\mp v_0}{\omega |x_{t_i}|})$. Denoting the auxiliary process as
$q_{\mathrm{aux}}(x_t,t)$, changing velocity direction at position $x_t$ at
time $t$, and taking $p_{\mathrm{aux}}(x,t)$ as the PDF of finding the
auxiliary process staying at $x$ at time $t$. Therefore we have
\begin{equation}
\label{supp_sec1_eq1}
q_{\mathrm{aux}}(x_t,t)=\int_{-\infty}^{\infty}\int_0^tq_{\mathrm{aux}}(x_{t-\tau},
t-\tau)\upsilon(x_t,x_{t-\tau},\tau)\phi(\tau)d\tau d x_{t-\tau}+p_0(x)\delta(t),
\end{equation}
where $\delta(\cdot)$ represents the Dirac $\delta$-function, $p_0(x)$ is the
initial PDF, $\upsilon(x_t,x_{t-\tau},\tau)=\frac{1}{2}\delta(x_t-A\cos(\omega
\tau+\varphi_{v_0}))+\frac{1}{2}\delta(x_t-A\cos(\omega\tau+\varphi_{-v_0}))$,
$\varphi_{\pm v_0}=\arctan\big(\frac{\mp v_0}{\omega|x_{t-\tau}|}$\big) for
$x_{t-\tau} \neq 0$, $\varphi_{\pm v_0}=\mp \frac{\pi}{2}$ when $x_{t-\tau}=0$.
According to probability theory, if we choose the initial distribution as
$p_0(x)=\delta(x)$, that is $x_{t-\tau}\equiv0$ when $t-\tau=0$, and the probability
distribution at a given point $t=\tau$ is zero, then the probability of $x_{t-\tau}
=0$ is also zero when $t-\tau\neq 0$. Thus, without loss of generality, in the
following we only need to consider $x_{t-\tau}\neq0$. Moreover, it can be verified
that
\begin{equation}
\label{upsilon_rw}
\begin{split}
\upsilon(x_t,x_{t-\tau},\tau)=&\frac{1}{2}\delta\left(x_t-\sqrt{x_{t-\tau}^2
+\frac{v_0^2}{\omega^2}}\cos\left(\omega \tau+\arctan\left(\frac{-v_0}{\omega
x_{t-\tau}}\right)\right)\right)  \\
&+\frac{1}{2}\delta\left(x_t-\sqrt{x_{t-\tau}^2+\frac{v_0^2}{\omega^2}}\cos\left(
\omega \tau+\arctan\left(\frac{v_0}{\omega x_{t-\tau}}\right)\right)\right).
\end{split}
\end{equation}
Consider the property of the $\delta$-function
\begin{equation}
\label{prop_delta_function}
\delta(g(x))=\sum_{i}\frac{\delta(x-x_i)}{|g'(x_i)|},
\end{equation}
where $x_i$ is the root of $g(x)=0$ and the sum in Eq. (\ref{prop_delta_function})
extends over all roots. In order to utilize Eq. (\ref{prop_delta_function}) to
simplify $\upsilon(x,\tau)$, we need to solve the following equations first
\begin{align}
x_t-\sqrt{x_{t-\tau}^2+\frac{v_0^2}{\omega^2}}\cos\left(\omega\tau+\arctan\left(
\frac{-v_0}{\omega x_{t-\tau}}\right)\right) &=0
\label{eqts_simp_epsilon_1},\\
x_t-\sqrt{x_{t-\tau}^2+\frac{v_0^2}{\omega^2}}\cos\left(\omega \tau+\arctan\left(
\frac{v_0}{\omega x_{t-\tau}}\right)\right) &=0
\label{eqts_simp_epsilon_2}.
\end{align}
From Eq. (\ref{eqts_simp_epsilon_1}), there exists
\begin{equation*}
x_t=\frac{\sqrt{\omega^2 x_{t-\tau}^2+v_0^2}}{\omega}\left[\cos(\omega\tau)\frac{
\omega|x_{t-\tau}|}{\sqrt{\omega^2 x_{t-\tau}^2+v_0^2}} +\sin(\omega \tau)\frac{
v_0|x_{t-\tau}|}{x_{t-\tau}\sqrt{\omega^2 x_{t-\tau}^2+v_0^2}}\right],
\end{equation*}
which can be equivalently written as
\begin{equation}
\label{sol_eqts_simp_epsilon_1}
x_{t-\tau}=\begin{cases}\frac{x_t}{\cos(\omega\tau)}-\frac{v_0}{\omega}\tan(\omega
\tau), & \mbox{if $x_{t-\tau}>0$};\\
-\frac{x_t}{\cos(\omega \tau)}-\frac{v_0}{\omega}\tan(\omega\tau), & \mbox{if
$x_{t-\tau}<0$}.
\end{cases}
\end{equation}
Moreover, it can be obtained that $|g'(x_{t-\tau})|=|\cos(\omega\tau)|$, where here
$g(y)=x_t-\sqrt{y+\frac{v_0^2}{\omega^2}}\cos\left(\omega\tau+\arctan\left(\frac{-
v_0}{\omega y}\right)\right)$. Therefore we have
\begin{equation}
\label{delta_func_1}
\begin{split}
\delta(x_t-A\cos(\omega\tau+\varphi_{v_0}))=&\frac{\Theta(x_{t-\tau})}{|\cos(\omega
\tau)|}\delta\left(x_{t-\tau}-\frac{x_t}{\cos(\omega \tau)}+\frac{v_0}{\omega}\tan(
\omega\tau)\right)\\
&+\frac{\Theta(-x_{t-\tau})}{|\cos(\omega \tau)|}\delta\left(x_{t-\tau}+\frac{x_t}{
\cos(\omega\tau)}+\frac{v_0}{\omega}\tan(\omega\tau)\right),
\end{split}
\end{equation}
where $\Theta(x)=1$ when $x>0$, otherwise $\Theta(x)=0$. Similarly we have
\begin{equation}
\label{delta_func_2}
\begin{split}
\delta(x_t-A\cos(\omega\tau+\varphi_{-v_0}))=&\frac{\Theta(x_{t-\tau})}{|\cos(
\omega\tau)|}\delta\left(x_{t-\tau}-\frac{x_t}{\cos(\omega\tau)}-\frac{v_0}{\omega}
\tan(\omega\tau)\right)\\
&+\frac{\Theta(-x_{t-\tau})}{|\cos(\omega\tau)|}\delta\left(x_{t-\tau}+\frac{x_t}{
\cos(\omega\tau)}-\frac{v_0}{\omega}\tan(\omega\tau)\right).
\end{split}
\end{equation}
Combining the definition of $\upsilon(x_t,x_{t-\tau},\tau)$, Eq. (\ref{delta_func_1})
and Eq. (\ref{delta_func_2}), then Eq. \eqref{supp_sec1_eq1} can be rewritten as
\begin{equation}
\label{final_sec_1}
\begin{split}
q_{\mathrm{aux}}(x_t,t)=&\frac{1}{2}\int_0^t\frac{1}{|\cos(\omega \tau)|}q_{
\mathrm{aux}}\left(\frac{x_t}{\cos(\omega\tau)}-\frac{v_0}{\omega}\tan(\omega
\tau),t-\tau\right)\Theta\left(\frac{x_t}{\cos(\omega \tau)}-\frac{v_0}{\omega}
\tan(\omega\tau)\right)\phi(\tau)d\tau\\
&\hspace*{-0.8cm}+\frac{1}{2}\int_0^t\frac{1}{|\cos(\omega \tau)|}q_{\mathrm{
aux}}\left(-\frac{x_t}{\cos(\omega\tau)}-\frac{v_0}{\omega}\tan(\omega\tau),t-
\tau\right)\Theta\left(\frac{x_t}{\cos(\omega \tau)}+\frac{v_0}{\omega}\tan(
\omega\tau)\right)\phi(\tau)d\tau\\
&\hspace*{-0.8cm}+\frac{1}{2}\int_0^t\frac{1}{|\cos(\omega\tau)|}q_{\mathrm{
aux}}\left(\frac{x_t}{\cos(\omega\tau)}+\frac{v_0}{\omega}\tan(\omega\tau),t-
\tau\right)\Theta\left(\frac{x_t}{\cos(\omega\tau)}+\frac{v_0}{\omega}\tan(
\omega\tau)\right)\phi(\tau)d\tau\\
&\hspace*{-0.8cm}+\frac{1}{2}\int_0^t\frac{1}{|\cos(\omega\tau)|}q_{\mathrm{
aux}}\left(-\frac{x_t}{\cos(\omega\tau)}+\frac{v_0}{\omega}\tan(\omega\tau),t-
\tau\right)\Theta\left(\frac{x_t}{\cos(\omega\tau)}-\frac{v_0}{\omega}\tan(
\omega\tau)\right)\phi(\tau)d\tau+p_0(x)\\
&\hspace*{-0.8cm}+\delta(t).
\end{split}
\end{equation}

\section{Derivation of the recursive relations of $\{\hat{T}_n(s)\}$ and $\{\hat{
\tilde{T}}_n(s)\}$}
\label{supp_sec2}

After obtaining Eq. (\ref{final_sec_1}) and assuming that $q_{{\rm aux}}(x,t)=\sum_{n=0}^{
\infty}H_n(x)e^{-x^2}T_n(t)$, where $H_n(x)$ are Hermite polynomials \cite{hermit_11,
hermit_intro1} and $T_n(t)$ are functions to be determined. Then we find
\begin{equation}
\label{q_Hermit}
\begin{split}
\sum_{n=0}^{\infty}H_n(x)e^{-x^2}T_n(t)=&\frac{1}{2}\int_0^t\frac{1}{|\cos(\omega
\tau)|}\sum_{n=0}^{\infty}H_n(x^-)e^{-(x^-)^2}T_n(t-\tau)\Theta(x^-)\phi(\tau)d\tau\\
&+\frac{1}{2}\int_0^t\frac{1}{|\cos(\omega\tau)|}\sum_{n=0}^{\infty}H_n(-x^+)e^{
-(x^+)^2}T_n(t-\tau)\Theta(x^+)\phi(\tau)d\tau\\
&+\frac{1}{2}\int_0^t\frac{1}{|\cos(\omega\tau)|}\sum_{n=0}^{\infty}H_n(x^+)e^{
-(x^+)^2}T_n(t-\tau)\Theta(x^+)\phi(\tau)d\tau\\
&+\frac{1}{2}\int_0^t\frac{1}{|\cos(\omega\tau)|}\sum_{n=0}^{\infty}H_n(-x^-)e^{
-(x^-)^2}T_n(t-\tau)\Theta(x^-)\phi(\tau)d\tau+p_0(x)\delta(t),
\end{split}
\end{equation}
where $x^\pm=\frac{x}{\cos(\omega \tau)}\pm\frac{v_0}{\omega}\tan(\omega \tau)$.
Multiplying by $H_m(x)$, $m=0,1,\ldots$ on both sides of Eq.~(\ref{q_Hermit}),
integrating over $(-\infty,\infty)$ with respect to $x$, and changing variables,
yield
\begin{equation}
\label{q_Hermit_2}
\begin{split}
&\sum_{n=0}^{\infty}\int_{-\infty}^{\infty}H_n(x)H_m(x)e^{-x^2}T_n(t)\\
&=\frac{1}{2}\int_0^{\infty}\int_0^t\sum_{n=0}^{\infty}H_n(y)H_m\left(\cos(\omega
\tau)y+\frac{v_0}{\omega}\sin(\omega\tau)\right)e^{-y^2}T_n(t-\tau)\phi(\tau)d\tau\\
&+\frac{1}{2}\int_{-\infty}^0\int_0^t\sum_{n=0}^{\infty}H_n(y)H_m\left(-\cos(\omega
\tau)y-\frac{v_0}{\omega}\sin(\omega\tau)\right)e^{-y^2}T_n(t-\tau)\phi(\tau)d\tau\\
&+\frac{1}{2}\int_0^{\infty}\int_0^t\sum_{n=0}^{\infty}H_n(y)H_m\left(\cos(\omega
\tau)y-\frac{v_0}{\omega}\sin(\omega\tau)\right)e^{-y^2}T_n(t-\tau)\phi(\tau)d\tau\\
&+\frac{1}{2}\int_{-\infty}^0\int_0^t\sum_{n=0}^{\infty} H_n(y)H_m\left(-\cos(\omega
\tau)y+\frac{v_0}{\omega}\sin(\omega\tau)\right)e^{-y^2}T_n(t-\tau)\phi(\tau)d\tau
+H_m(0)\delta(t).
\end{split}
\end{equation}

First we consider even $m$. For Hermite polynomials the symmetry relation $H_m(x)=
H_m(-x)$ holds for even $m$, thus the right hand side of Eq.~(\ref{q_Hermit_2}) can
be rewritten as
\begin{equation*}
\begin{split}
&\frac{1}{2}\int_{-\infty}^{\infty}\int_0^t\sum_{n=0}^{\infty}H_n(y)H_m\left(\cos(
\omega\tau)y+\frac{v_0}{\omega}\sin(\omega\tau)\right)e^{-y^2}T_n(t-\tau)\phi(\tau)
d\tau\\
&+\frac{1}{2}\int_{-\infty}^{\infty}\int_0^t\sum_{n=0}^{\infty} H_n(y)H_m\left(\cos(
\omega\tau)y-\frac{v_0}{\omega}\sin(\omega\tau)\right)e^{-y^2}T_n(t-\tau)\phi(\tau)d
\tau+H_m(0)\delta(t).
\end{split}
\end{equation*}
With the properties of the Hermite polynomials shown in Eqs. \eqref{hermit_prop_sum}, \eqref{hermit_prop_time},
we find
\begin{equation*}
\begin{split}
&\sqrt{\pi}2^mm!T_m(t)-(-2)^{\frac{m}{2}}(m-1)!!\delta(t)\\
&=\frac{1}{2}\sum_{j=0}^m\sum^{\lfloor\frac{j}{2}\rfloor}_{i=0}\int_0^t\frac{m!
\sqrt{\pi}2^{j-2i}}{i!(m-j)!}\left(\frac{2 v_0}{\omega}\right)\left[\sin^{m-j}(
\omega\tau)+(-\sin(\omega\tau))^{m-k}\right]\cos^{k-2i}(\omega\tau)(-\sin^2(
\omega\tau))^iT_{k-2i}(t-\tau)\phi(\tau)d\tau.
\end{split}
\end{equation*}
Taking the Laplace transform with respect to $t$, we finally obtain
\begin{equation}
\label{recur_T}
\begin{split}
&\sqrt{\pi}2^mm!\hat{T}_m(s)-(-2)^{\frac{m}{2}}(m-1)!!\\
&=\frac{1}{2}\sum_{j=0}^m\sum^{\lfloor\frac{j}{2}\rfloor}_{i=0}\frac{m!\sqrt{\pi}
2^{j-2i}}{i!(m-j)!}\left(\frac{2 v_0}{\omega}\right)\mathscr{L}\left\{\left[\sin^{
m-j}(\omega\tau)+(-\sin(\omega\tau))^{m-k}\right]\cos^{k-2i}(\omega\tau)(-\sin^2(
\omega\tau))^i\phi(\tau)\right\}\hat{T}_{k-2i}(s).
\end{split}
\end{equation}
Similarly we assume that $p_{{\rm aux}}(x,t)=\sum_{n=0}^{\infty}H_n(x)e^{-x^2}\tilde{T}_n(t)$,
and it then follows that
\begin{equation}
\label{recur_TT}
\begin{split}
&\sqrt{\pi}2^m m! \hat{\tilde{T}}_m(s)\\
&=\frac{1}{2}\sum_{j=0}^{m}\sum^{\lfloor\frac{j}{2}\rfloor}_{i=0}\frac{m!\sqrt{\pi}
2^{j-2i}}{i!(m-j)!}\left(\frac{2v_0}{\omega}\right)\mathscr{L}\left\{\left[\sin^{m
-j}(\omega\tau)+(-\sin(\omega\tau))^{m-k}\right]\cos^{k-2i}(\omega\tau)(-\sin^2(
\omega\tau))^{i}\Psi(\tau)\right\}\hat{T}_{k-2i}(s),
\end{split}
\end{equation}
where $\Psi(\tau)=\int_{\tau}^{\infty}\phi(\tau')d\tau'$ is the survival probability.

For odd $m$, the Hermite polynomials satisfy the antisymmetric relation $H_m(x)=-
H_m(-x)$, therefore the right hand side of Eq.~(\ref{q_Hermit_2}) can be rewritten
as
\begin{equation}
\label{rhs_s11}
\begin{split}
&\frac{1}{2}\sum_{n=0}^{\infty}\sum_{j=0}^{m}\sum_{i=0}^{\lfloor\frac{j}{2}\rfloor}
\int_0^t\frac{m!}{i!(j-2i)!(m-j)!}\left[\left(\frac{2 v_0}{\omega}\sin(\omega\tau)
\right)^{m-j}\cos^{j-2i}(\omega\tau)(-\sin^2(\omega\tau))^i\int_0^{\infty}H_{j-2i}
(y)H_n(y)e^{-y^2}dy\right.\\
&+\left(-\frac{2v_0}{\omega}\sin(\omega\tau)\right)^{m-j}(-\cos(\omega\tau))^{j-2i}
(-\sin^2(\omega\tau))^i\int_{-\infty}^0H_{j-2i}(y)H_n(y)e^{-y^2}dy\\
&+\left(-\frac{2v_0}{\omega}\sin(\omega\tau)\right)^{m-j}\cos^{j-2i}(\omega\tau)(-
\sin^2(\omega\tau))^i\int_0^{\infty}H_{j-2i}(y)H_n(y)e^{-y^2}dy\\
&\left.+\left(\frac{2v_0}{\omega}\sin(\omega\tau)\right)^{m-j}(-\cos(\omega\tau))^{
j-2i}(-\sin^2(\omega\tau))^i\int_{-\infty}^0H_{j-2i}(y)H_n(y)e^{-y^2}dy\right]\phi(
\tau)T_n(t-\tau)d\tau,
\end{split}
\end{equation}
which indicates that when $m$ is odd, $j$ in Eq.~(\ref{rhs_s11}) must be odd as well,
otherwise expression Eq. (\ref{rhs_s11}) equals zero. Therefore Eq.~(\ref{rhs_s11}) can
be further rewritten as
\begin{equation}
\label{rw_rhs_s11}
\begin{split}
&\sum_{n=0}^{\infty}\sum_{j=0}^{\lfloor\frac{m}{2}\rfloor}\sum_{i=0}^j\int_0^t
\frac{m!}{i!(2j+1-2i)!(m-2j-1)!}\left(\frac{2 v_0}{\omega}\sin(\omega \tau)\right)^{
m-2j-1}\cos^{2j+1-2i}(\omega\tau)(-\sin^2(\omega\tau))^i\\
&\left[\int_0^{\infty}H_{2j+1-2i}(y)H_n(y)e^{-y^2}dy-\int_{-\infty}^0H_{2j+1-2i}(y)
H_n(y)e^{-y^2}dy\right]T_n(t-\tau)\phi(\tau)d\tau,
\end{split}
\end{equation}
which indicates that when $n$ is odd, $H_{2j+1-2i}(y)H_n(y)e^{-y^2}$ is an even
function, further $\int_0^{\infty}H_{2j+1-2i}(y)H_n(y)e^{-y^2}dy-\int_{-\infty}^0
H_{2j+1-2i}(y)H_n(y)e^{-y^2}dy=0$, i.e., Eq.~(\ref{rw_rhs_s11}) is zero. Therefore,
odd terms of $T_n(t)$ disappear and there exists
\begin{equation}
\label{odd_m_T}
\begin{split}
\sqrt{\pi}2^mm!T_m(t)=&2\sum_{n=0}^{\infty}\sum_{j=0}^{\lfloor\frac{m}{2}\rfloor}
\sum_{i=0}^j\frac{m!}{i!(2j+1-2i)!(m-2j-1)!}\int_0^t\left(\frac{2 v_0}{\omega}\sin(
\omega\tau)\right)^{m-2j-1}\cos^{2j+1-2i}(\omega\tau)(-\sin^2(\omega\tau))^i\\
&\times\int_0^{\infty}H_{2j+1-2i}(y)H_{2n}(y)e^{-y^2}dyT_{2n}(t-\tau)\phi(\tau)d\tau.
\end{split}
\end{equation}
We now use the following property of the Hermite polynomials \cite{S and I}
\begin{equation*}
\int_0^{\infty}H_{2j+1-2i}(y)H_{2n}(y)e^{-y^2}dy=\frac{\sqrt{\pi}~_2F_1\left(
-(2j+1-2i),-2n;1-\frac{2j+1-2i}{2}-n;\frac{1}{2}\right)}{2^{1-(2j+1-2i)-2n}
\Gamma\left(1- \frac{2j+1-2i}{2}-n\right)},
\end{equation*}
where $_2F_1(a,b;c;d)$ is the hypergeometric function defined as
\begin{equation*}
_2F_1(a,b;c;d)=\sum_{n=0}^{\infty}\frac{(a)_n(b)_n}{(c)_n}\frac{z^n}{n!}.
\end{equation*}
Here $(d)_n$ represents the Pochhammer symbol,
\begin{equation*}
(d)_n=\begin{cases}1,&\mbox{if $n=0$};\\
d(d+1)\cdots(d+n-1),&\mbox{if $n>0$}.
\end{cases}
\end{equation*}
After taking the Laplace transform with respect to $t$ of Eq.~(\ref{odd_m_T}) we
have
\begin{equation}
\label{odd_m_hatT}
\begin{split}
\sqrt{\pi}2^mm!\hat{T}_m(s)=&2\sum_{n=0}^{\infty}\sum_{j=0}^{\lfloor\frac{m}{2}
\rfloor}\sum_{i=0}^j\frac{m!\sqrt{\pi}~_2F_1\left(-(2j+1-2i),-2n;\frac{1}{2}+i
-j-n;\frac{1}{2}\right)}{i!(2j+1-2i)!(m-2j-1)!\Gamma\left(\frac{1}{2}+i-j-n\right)
2^{2(i-j-n)}}\left(\frac{2 v_0}{\omega}\right)^{m-2j-1}\\
&\times\mathscr{L}\left\{\sin^{m-2j-1}(\omega\tau)\cos^{2j+1-2i}(\omega\tau)(-
\sin^2(\omega\tau))^i\phi(\tau)\right\}\hat{T}_{2n}(s).
\end{split}
\end{equation}
Similarly, for odd $m$ we obtain the relations of $\hat{\tilde{T}}_m(s)$ and
$\hat{T}_{2n}(s)$,
\begin{equation}
\label{odd_m_TT_L}
\begin{split}
\sqrt{\pi}2^mm!\hat{\tilde{T}}_m(s)=&2\sum_{n=0}^{\infty}\sum_{j=0}^{\lfloor
\frac{m}{2}\rfloor}\sum_{i=0}^j\frac{m!\sqrt{\pi}~_2F_1\left(-(2j+1-2i),-2n;
\frac{1}{2}+i-j-n;\frac{1}{2}\right)}{i!(2j+1-2i)!(m-2j-1)!\Gamma\left(\frac{
1}{2}+i-j-n\right)2^{2(i-j-n)}}\left(\frac{2v_0}{\omega}\right)^{m-2j-1}\\
&\times\mathscr{L}\left\{\sin^{m-2j-1}(\omega\tau)\cos^{2j+1-2i}(\omega\tau)(
-\sin^2(\omega\tau))^i\Psi(\tau)\right\}\hat{T}_{2n}(s).
\end{split}
\end{equation}

\section{Approximate stationary distribution for L\'evy walks in harmonic
potential with free and reflecting boundary conditions}
\label{supp_sec3}

In this section, we mainly provide the approximate results for the stationary
PDF for LWs in a harmonic potential when the duration of individual walk steps
$\tau$ follows the exponential density $e^{-\tau}$. For simplicity of calculations
we take $v_0=\omega=1$. First we provide results for free boundary conditions. In
this case the odd terms of $\{T_n(t)\}$ and $\{\tilde{T}_n(t)\}$ vanish. Therefore
it is sufficient to consider the even terms, which can be represented as the
recurrence formulas as Eq. \eqref{fb_T_laplace} and Eq. \eqref{fb_tildeT_laplace}. The behaviors of $\tilde{T}_0(t),
\tilde{T}_2(t),\ldots,\tilde{T}_{12}(t)$ for $t\to\infty$ are then
\begin{equation*}
\begin{split}
&\lim_{t\to\infty}\tilde{T}_0(t)=\frac{1}{\sqrt{\pi}};\\
&\lim_{t\to\infty}\tilde{T}_2(t)=\frac{1}{4\sqrt{\pi}};\\
&\lim_{t\to\infty}\tilde{T}_4(t)=-\frac{5}{352\sqrt{\pi}};\\
&\lim_{t\to\infty}\tilde{T}_6(t)=\frac{63}{2721928\sqrt{\pi}};\\
&\lim_{t\to\infty}\tilde{T}_8(t)=\frac{146606719}{5558830238720\sqrt{\pi}};\\
&\lim_{t\to\infty}\tilde{T}_{10}(t)=-\frac{39362159928909}{30088502786329292800
\sqrt{\pi}};\\
&\lim_{t\to\infty}\tilde{T}_{12}(t)=\frac{198428708025937940281}{
4895922461868213486986035200\sqrt{\pi}}.\\
\end{split}
\end{equation*}
The approximate stationary distribution $p^{\mathrm{st}}(x)\approx\sum_{n=0}^6
\lim_{t\to\infty}\tilde{T}_{2n}(t)H_{2n}(x)e^{-x^2}$ is shown in Fig. 1 in the
main text.

For the case of \emph{reflecting boundary}, the behaviors of $\tilde{T}_0(t),\tilde{T}_2(t),\ldots,\tilde{T}_{12}(t)$ are obtained from  Eqs. \eqref{recur_T} and \eqref{recur_TT} as follows
\begin{equation*}
\begin{split}
&\lim_{t\to\infty}\tilde{T}_0(t)=\frac{1}{\sqrt{\pi}};\\
&\lim_{t\to\infty}\tilde{T}_2(t)=\frac{1}{4\sqrt{\pi}};\\
&\lim_{t\to\infty}\tilde{T}_4(t)=-\frac{5}{352\sqrt{\pi}};\\
&\lim_{t\to\infty}\tilde{T}_6(t)=-\frac{7}{429440\sqrt{\pi}};\\
&\lim_{t\to\infty}\tilde{T}_8(t)=\frac{112707}{4385097728\sqrt{\pi}};\\
&\lim_{t\to\infty}\tilde{T}_{10}(t)=-\frac{151918210141}{118676969381273600
\sqrt{\pi}};\\
&\lim_{t\to\infty}\tilde{T}_{12}(t)=\frac{154859416018475893}{
3862161199753787756052480\sqrt{\pi}}.\\
\end{split}
\end{equation*}
For the odd terms, due to the involved terms we only use $T_0(t),T_2(t),\ldots,
T_8(t)$ for their approximate calculations, then utilizing Eq. \eqref{odd_m_TT_L} leads us to the results
\begin{equation*}
\begin{split}
&\lim_{t\to\infty}\tilde{T}_1(t)\approx\frac{2124385847}{2740686080 \pi};\\
&\lim_{t\to\infty}\tilde{T}_3(t)\approx\frac{6463601801}{164441164800 \pi};\\
&\lim_{t\to\infty}\tilde{T}_5(t)\approx-\frac{18361300219}{3420376227840 \pi};\\
&\lim_{t\to\infty}\tilde{T}_7(t)\approx-\frac{687695174759}{3990438932480000 \pi};\\
&\lim_{t\to\infty}\tilde{T}_9(t)\approx\frac{70824923727473}{
70678654372085760000\pi};\\
&\lim_{t\to\infty}\tilde{T}_{11}(t)\approx-\frac{1483140637545367}{
17245591666788925440000\pi}.\\
\end{split}
\end{equation*}
Finally $p_{\mathrm{aux}}^{\mathrm{st}}(x)\approx\sum_{n=0}^{12}\lim_{t\to\infty}
\tilde{T}_n(t)H_n(x)e^{-x^2}$, showing good convergence for the involved number of
terms. The stationary PDF for the case of a reflecting boundary is then $p^{\mathrm{
st}}_{\mathrm{rb}}(x)=p_{\mathrm{aux}}^{\mathrm{st}}(x)+p_{\mathrm{aux}}^{\mathrm{
st}}(-x)$ for $x\geq 0$. The corresponding simulation results are shown in Fig.~1
of the main text.


\begin{thebibliography}{2019}

\bibitem{bouchaud} J.-P. Bouchaud and A. Georges, Phys. Rep. \textbf{195}, 127
(1990).

\bibitem{pccp} R. Metzler, J.-H. Jeon, A. G. Cherstvy and E. Barkai, Phys. Chem. Chem. Phys. \textbf{16}, 24128 (2014).

\bibitem{franosch} F. H\"ofling and T. Franosch, Rep. Progr. Phys.
\textbf{76}, 046602 (2013); K. N{\o}rregaard, R. Metzler, C. Ritter,
K. Berg-S{\o}rensen, and L. Oddershede, Chem. Rev. \textbf{117}, 4342 (2017).

\bibitem{scher} H. Scher and E. W. Montroll, Phys. Rev. B \textbf{12}, 2455
(1975).

\bibitem{weiss} J. Szymanski and M. Weiss, Phys. Rev. Lett. \textbf{103},
038102 (2009); W. Pan, L. Filobelo, N. D. Q. Pham, O. Galkin, V. V. Uzunova, and
P. G. Vekilov Phys. Rev. Lett. \textbf{102}, 058101 (2009);
J.-H. Jeon, N. Leijnse, L. B. Oddershede, and R. Metzler, New
J. Phys. \textbf{15}, 045011 (2013).

\bibitem{schwille} P. Schwille, U. Haupts, S. Maiti, and W. W. Webb, Biophys. J.
\textbf{77}, 2251 (1999); M. Weiss, H. Hashimoto, and T. Nilsson, Biophys. J.
\textbf{84}, 4043 (2003); S. Gupta, J. U. de Mel, R. M. Perera, P. Zolnierczuk,
M. Bleuel, A. Faraone, and G. J. Schneider, J. Phys. Chem. Lett. \textbf{9},
2956 (2018); W. He, H. Song, Y. Su, L. Geng, B. J. Ackerson, H. B. Peng, and P.
Tong, Nat. Comm. \textbf{7}, 11701 (2016).

\bibitem{weigel} A. V. Weigel, B. Simon, M. M. Tamkun and D. Krapf,
Proc. Natl. Acad. Sci. U. S. A. \textbf{108}, 6438 (2011);
C. Manzo, J. A. Torreno-Pina, P. Massignan, G. J. Lapeyre, Jr.,
M. Lewenstein, and M. F. Garcia Parajo, Phys. Rev. X \textbf{5}, 011021 (2015).

\bibitem{ilpo} G. R. Kneller, K. Baczynski, and M. Pasienkewicz-Gierula,
J. Chem. Phys. \textbf{135}, 141105 (2011); J.-H. Jeon, H. M. Monne, M.
Javanainen, and R. Metzler, Phys. Rev. Lett. \textbf{109}, 188103 (2012);
J.-H. Jeon, M. Javanainen, H. Martinez-Seara, R. Metzler, and I. Vattulainen,
Phys. Rev. X \textbf{6}, 021006 (2016).

\bibitem{lene} S. C. Weber, A. J. Spakowitz, and J. A. Theriot, Phys. Rev. Lett.
\textbf{104}, 238102 (2010); I. Golding and E. C. Cox, Phys. Rev. Lett. \textbf{96},
098102 (2006); I. Bronstein, Y. Israel, E. Kepten, S. Mai, Y. Shav-Tal,
E. Barkai and Y. Garini, Phys. Rev. Lett. \textbf{103}, 018102 (2009);
J.-H. Jeon, V. Tejedor, S. Burov, E. Barkai, C. Selhuber-Unkel,
K. Berg-S{\o}rensen, L. Oddershede, and R. Metzler, Phys. Rev. Lett. \textbf{106},
048103 (2011).

\bibitem{tabei} S. M. Tabei, S. Burov, H. Y. Kim, A. Kuznetsov, T. Huynh, J.
Jureller, L. H. Philipson, A. R. Dinner, and N. F. Scherer, Proc. Natl. Acad. Sci.
U.S.A. \textbf{110}, 4911 (2013).

\bibitem{grl} Y. Edery, H. Scher, A. Guadagnini, and B. Berkowitz, Water Res. Res.
\textbf{50}, 1490 (2014); N. Goeppert, N. Goldscheider, and B. Berkowitz, Water Res.
Res., in press, DOI:10.1016/j.watres.2020.115755.

\bibitem{roberts} D. Robert, T. H. Nguyen, F. Gallet, and C. Wilhelm, PLoS ONE
\textbf{4}, e10046 (2010); A. Caspi, R. Granek, and M. Elbaum. Phys. Rev. Lett.
\textbf{85}, 5655 (2000).

\bibitem{christine} J. F. Reverey, J.-H. Jeon, M. Leippe, R. Metzler, and C.
Selhuber-Unkel, Sci. Rep. \textbf{5}, 11690 (2015).

\bibitem{seisenhuber} G. Seisenberger, M. U. Ried, T. Endre{\ss}, H. B{\"u}ning,
M. Hallek, and C. Br{\"a}uchle, Science \textbf{294}, 1929 (2001).

\bibitem{igorturb} G. Boffetta and I. M. Sokolov, Phys. Rev. Lett. \textbf{88},
094501 (2002).

\bibitem{montroll} E. W. Montroll and G. H. Weiss, J. Math. Phys. \textbf{10},
753 (1969).

\bibitem{report} R. Metzler and J. Klafter, Phys. Rep. \textbf{339}, 1 (2000).

\bibitem{mebakla} R. Metzler, E. Barkai, and J. Klafter, Phys. Rev. Lett.
\textbf{82}, 3563 (1999); Europhys. Lett. \textbf{46}, 431 (1999).

\bibitem{fogedby} H. C. Fogedby, Phys. Rev. E \textbf{50}, 1657 (1994); \emph{ibid.}
\textbf{58}, 1690 (1998); Phys. Rev. Lett. \textbf{73}, 2517 (1994).

\bibitem{jespersen} S. Jespersen, R. Metzler, and H. C. Fogedby, Phys. Rev. E
\textbf{59} 2736 (1999).

\bibitem{ghandi} G.M. Viswanathan, V. Afanasyev, S.V. Buldyrev, E.J. Murphy,
P.A. Prince, and H.E. Stanley, Nature \textbf{381}, 413 (1996);
G. M Viswanathan, M. G. E. da Luz, E. P. Raposo, and H. E. Stanley,
The Physics of Foraging (Cambridge University Press, Cambridge, 2011).

\bibitem{vladimir} V. V. Palyulin, A. V. Chechkin, and R. Metzler,
Proc. Natl. Acad. Sci. USA \textbf{111}, 2931 (2014).

\bibitem{chechkin} A. V. Chechkin, J. Klafter, V. Yu. Gonchar, R. Metzler, and L. V.
Tanatarov, Phys. Rev. E \textbf{67}, 010102(R) (2003); A. V. Chechkin, V. Yu. Gonchar,
J. Klafter, and R. Metzler, Phys. Rev. E \textbf{72} 010101(R) (2005).

\bibitem{wang} M. F. Shlesinger, J. Klafter and Y. M. Wong, J. Stat. Phys.
\textbf{27}, 499 (1982); M. F. Shlesinger and J. Klafter, Phys. Rev. Lett.
\textbf{54}, 2551 (1985).

\bibitem{zaburdaev} V. Zaburdaev, S. Denisov, and J. Klafter, Rev. Mod. Phys.
\textbf{87}, 483 (2015).

\bibitem{dhar} P. Cipriani, S. Denisov, and A. Politi, Phys. Rev. Lett.
\textbf{94}, 244301 (2005); A. Dhar, K. Saito, B. Derrida, Phys. Rev. E
\textbf{87}, 010103(R) (2013).

\bibitem{eliyossi} E. Barkai, V. Fleurov, and J. Klafter, Phys. Rev. E
\textbf{61}, 1164 (2000).

\bibitem{light} R. Patel and R. Mehta, J. Nanophot. \textbf{6}, 069503 (2012);
P. Barthelemy, J. Bertolotti, and D. S. Wiersma, Nature \textbf{453}, 495 (2008).

\bibitem{michael} M. A. Lomholt, T. Koren, R. Metzler, and J. Klafter,
Proc. Natl. Acad. Sci. USA \textbf{105}, 11055 (2008).

\bibitem{sims} D. W. Sims, N. E. Humphries, N. Hu, V. Medan, and J. Berni, eLife
\textbf{8}, e50316 (2019).

\bibitem{vladimir1} V. V. Palyulin, G. Blackburn, M. A. Lomholt, N. Watkins, R.
Metzler, R. Klages, and A. V. Chechkin, New J. Phys. \textbf{21}, 103028 (2019).

\bibitem{abe} M. S. Abe, E-print bioRxiv:2020.01.27.920801.

\bibitem{jae} M. S. Song, H. C. Moon, J.-H. Jeonm, and H. Y. Park, Nature Comm.
\textbf{9}, 344 (2018); K. J. Chen, B. Wang, and S. Granick, Nature Mater.
\textbf{14}, 589 (2015).

\bibitem{cancer} S. Huda et al., Nature Comm. \textbf{9}, 4539 (2018).

\bibitem{hunt} D. A. Raichlen, B. M. Wood, A. D. Gordon, A. Z. Mabulla, F. W.
Marlowe, and H. Pontzer, Proc. Natl. Acad. Sci. USA \textbf{111}, 728 (2014).

\bibitem{pedestrian} H. Murakami, C. Feliciani, and K. Nishinari, J. Roy. Soc.
Interface \textbf{16}, 20180939 (2019).

\bibitem{robot} V. Fioriti, F. Fratichini, S Chiesa, and C. Moriconi, Int. J. Adv.
Robot. Syst. \textbf{12}, 98 (2015); Y. Katada, A. Nishiguchi, K. Moriwaki, and R.
Qatakabe, Artif. Life Robot. \textbf{21}, 295 (2016).

\bibitem{brockmann} D. Brockmann, L. Hufnagel, and T. Geisel, Nature \textbf{439},
462 (2006); A. Reynolds, E. Ceccon, C. Baldauf, T. K. Medeiros, and O. Miramontes,
PLoS ONE \textbf{13}, e0199099 (2018).

\bibitem{havlin} B. Gross, Z. Zheng, S. Liu, X. Chen, A. Sela, J. Li, D. Li, and
S. Havlin, E-print arXiv:2003.08382.

\bibitem{froemberg} D. Froemberg and E. Barkai, Phys. Rev. E \textbf{87}, 030104(R)
(2013); Euro. Phys. J. B \textbf{86}, 331 (2013).

\bibitem{godec} A. Godec and R. Metzler, Phys. Rev. Lett. \textbf{110}, 020603 (2013);
Phys. Rev. E \textbf{88}, 012116 (2013).

\bibitem{rebenstok} A. Rebenshtok, S. Denisov, P. H\"{a}nggi and E. Barkai, Phys.
Rev. Lett. \textbf{112}, 110601 (2014).

\bibitem{sokolov} I. M. Sokolov and R. Metzler, Phys. Rev. E \textbf{67}, 010101(R)
(2003).

\bibitem{friedrich} R. Friedrich, F. Jenko, A. Baule and S. Eule, Phys. Rev. Lett.
\textbf{96}, 230601 (2006); Phys. Rev. E \textbf{74}, 041103 (2006).

\bibitem{xu2018} P. B. Xu and W. H. Deng, J. Stat. Phys. \textbf{173}, 1598 (2018).

\bibitem{hermit_intro} M. Abramowitz and I. A. Stegun, \emph{Handbook of
Mathematical Functions} (Dover, New York, 1972).

\bibitem{hermit_1} P. B. Xu, W. H. Deng and T. Sandev, J. Phys. A: Math. Theor.
\textbf{53}(11), 115002 (2020).

\bibitem{supp} Supplemental material

\bibitem{vojta} A. H. O. Wada and T. Vojta, Phys. Rev. E \textbf{97}, 020102(R)
(2018); T. Guggenberger, G. Pagnini, T. Vojta, and R. Metzler, New J. Phys
\textbf{21}, 022002 (2019).

\bibitem{miron} A. Miron, Phys. Rev. E \textbf{100}, 012106 (2019);
Phys. Rev. Lett, \textbf{124}, 140601 (2020).

\end{thebibliography}

\begin{thebibliography}{1}

\bibitem{hermit_intro1} M. Abramowitz and I. A. Stegun, \emph{Handbook of
Mathematical Functions} (Dover, New York, 1972).

\bibitem{hermit_11} P. B. Xu, W. H. Deng and T. Sandev, J. Phys. A: Math. Theor.
\textbf{53}(11), 115002 (2020).

\bibitem{S and I}  A. P. Prudnikov, Y. A. Brychkov and O. I. Marichev,
\emph{Integral and Series} (Gordon \& Breach, New York, 1990).

\end{thebibliography}
\end{document}